\documentclass[a4paper,graphicx,12pt]{article}

\def\eq{\ =\ }
\def\mns{\ -\ }
\def\pls{\ +\ }
\def\be{\begin{equation}}
\def\ee{\end{equation}}

\usepackage{graphics}
\usepackage{epsfig}
\usepackage{graphicx}
\usepackage{color,soul}
\begin{document}
\begin{frontmatter}
\title{\bf Effects of structural distortion on electronic and optical properties of defect $CdGa_2X_4$ (X = S, Se, Te) Chalcopyrite Semiconductor}
\author[label1]{S.Mishra}
\author[label1]{B.Ganguli\corref{cor1}}
\ead {biplabg@nitrkl.ac.in}
\cortext[cor1]{corresponding author. Tel.: +91661 2462725; fax: +91661 2462999 }
\address[label1]{National Institute of Technology, Rourkela-769008, Odisha, India}
\begin{abstract}
We observe significant effects of structural distortion on electronic and optical properties of $CdGa_2X_4$ (X = S, Se, Te) defect chalcopyrite. The calculation is carried out within Density functional theory based tight binding linear muffin tin orbital (TB-LMTO) method. Structural parameters and band gap of $CdGa_2X_4$ agree well with the available experimental values within LDA limit. Change in band gap due to structural distortion is 3.63\%, 4.0\%, and 8.8\% respectively. We observe significant change in optical properties also due to this effect. Effects on optical properties come mainly from optical matrix elements. Our results of these response functions agree well with the available experimental values. 
\end{abstract}
\begin{keyword}
A. Defect chalcopyrite; A. Semiconductors; E. Density Functional Theory; E. TB-LMTO
 \end{keyword}                                                                                     
\end{frontmatter}
\section{Introduction}
$CdGa_2X_4$ (S, Se, Te) defect chalcopyrites are attractive materials because of their luminescent properties in the visible region \cite{1,2}. These compounds belong to group $II-III_2-VI_4$ defect chalcopyrites. Defect structure is due to vacancies at the cation sites. Vacancies are such that they do not break translational symmetry of the crystal. Defect structure makes such compounds porous. $CdGa_2X_4$ are not only important for device applications but they have attracted special attention of the physics community due their porisity nature. These vacancies can be occupied by suiatble elements to modify their physical properites without breaking symmetry. Therefore it is vital to know detailed structural, electronic and optical properties of these materials. Several experimental and theoretical studies are reported on these compounds \cite{3,4,5,6,7,8,9,10,11,12,13,14,15,16,17,18}.  Jiang et. al. \cite{9} studied band structure using first principle calculation. But no detailed electronic and optical properties were studied. Recently Ma et. al. \cite{14} and Zhao et. al. \cite{19} have reported  structural, electronic and optical properties of $CdGa_2X_4$ (X = S, Se, Te) using DFT based VASP code. It is well known that ternary chalcopyrites have some interesting structural anomalies \cite{20,21,22,23,24} relative to their binary analogous. These structural anomalies have significant effects on band gap and so on overall electronic and optical properties. It is well known that p-d hybridization does not have significant effect on band gap reduction in the group of $II-III_2-VI_4$ chalcopyrites \cite{21,23}. It is a prominant mechanism for band gap reduction in other groups of chalcopyrite semiconductors. Therefore the effect of structural distortion plays a major role in band gap increment/decrement in the group of $II-III_2-VI_4$ chalcopyrites. But there is no such study, either qualitative or quantitative, on these materials.\\  
Therefore the main motivation in this work is to study  effects of structural distortion on electronic and optical properties of $CdGa_2X_4$ (X = S, Se, Te). We consider here two types of structural distortions, the displacement of anions from their ideal position (‘u’) and structural deformation expressed by the ratio of the lattice parameters (c/2a). It has been established that structural distortion has prominent effects on band gap and electronic properties of  pure chalcopyrite \cite{22} and defect/substituted chalcopyrite semiconductors \cite{23,24}. In the present work we calculate quantitative change in band gap, electronic and optical properties due to structural distortion. Structural parameters like lattice parameters, tetragonal distortion, anion displacement parameters and bond lengths are calculated by structural relaxation using energy minimization proceedure. There have been no calculation so far of bulk modulus of $CdGa_2Te_4$. We calculate bulk modulus by fitting our data of energy versus volume to Murnaghan equation of states \cite{25}.  We carry out detailed electronic properties, like band structure, total density of states (TDOS) and partial density of states (PDOS) and then calculate effects of structural distortion on TDOS. To study effects of structural distortion on optical properties we calculate joint density of states (JDOS) and square of optical matrix elemnets (OME) separately. With these calculations we show how structural distortion have effects on optical response functions. Our calculations are based on density functional theory (DFT) \cite{26,27} first principle technique, tight binding linearized muffin-tin orbital (TB-LMTO) method \cite{28,29}. Basis functions in TB-LMTO method are localized. Therefore, very few such functions are required to represent highly localized d-orbitals. Hence the calculation is not only cost effective, it gives also accurate results.\\DFT requires a suitable choice of exchange and correlation functional for the electron-electron correlations. We have chosen such a functional within LDA(Local density approximation)  \cite{30}.  It is known that LDA underestimates band gap in semiconductors. Therefore it is necessary to use more advanced correlation functional like hybrid functional \cite{31}. But our motivation here is not to calculate band gap accurately. Purpose is to show  effects of structural distortion on electronic and optical properties. It is found LDA gives good structural properties of chalcopyrite semiconductors \cite{22,23,24,32} and calculate accurate band gap reduction due to p-d hybridization. This is because reduction does not depend significantly on the kind of functional used \cite{32}. We expect the same to be true also for reduction in band gap due to structural distortion.
\section{Methodology}
\subsection{Structural $\&$ Electronic Properties}
The ab initio method is based on Density Functional Theory of Kohn–Sham \cite{26,27}. The total electronic energy is a function of electron density which is calculated using variational principle. This requires self-consistent calculations. In practice the Kohn-Sham orbitals are usually expanded in terms of some chosen basis function. We have used the well-established TB-LMTO-ASA method, discussed in detail elsewhere \cite{28,29} for the choice of the basis function. To solve the one-electron Schrodinger-like equation self- consistently, we use the scalar-relativistic linear muffin-tin orbital method in the atomic sphere approximation (ASA) which includes the combined correction terms. The Hamiltonian and overlap matrices are evaluated over atomic spheres where the potential is spherically symmetric. Electron correlations are taken within LDA of DFT \cite{29}. We have used the von Barth–Hedin exchange \cite{30} with 512 {\bf k}-points in the irreducible part of the Brillouin zone. The basis of the TB-LMTO starts from the minimal set of muffin-tin orbitals of a KKR formalism and then linearizes it by expanding around a 'nodal' energy point represented by EKAP in band structure figures. The wave-function is then expanded in this basis. Basis functions up to $l_{max}$=3 for Cd and $l_{max}$=2 for both Ga $\&$ X (S, Se, Te) are used for the band structure calculation. Cd  4d, 5s $\&$ 5p, Ga 4s $\&$ 4p, S 3s $\&$ 3p, Se 4s $\&$ 4p, Te 5s $\&$ 5p and empty spheres (E, E1, E2, E3) 1s linearized muffin tin orbitals  are treated as valence states. The Cd 4f  Ga 4d, S 3d, Se 4d, Te 5d $\&$ 4f and empty spheres p and d states are downfolded for all the three systems in our calculation. These downfolded states are included in the tails of the valence LMTO's. These states do not contribute to the dimension of the Hamiltonian (H) and the overlap (O) matrices, but carry charge. This technique avoides the problem of 'ghost' bands. In our calculaion  the core states are treated relativistically, while the valence states were calculated scalar relativistically.\\
Defect chalcopyrite semiconductors are loosely packed crystals (low packing fraction). To overcome such unfavorable physical situation, it is necessary to introduce empty spheres at appropriate chosen interstitial sites for self-consistent calculation. This technique is incorporated in the advanced TB-LMTO-ASA method. We ensure proper overlap of muffin tin spheres for self-consistency and the percentage of overlaps are found to be 8.64$\%$, 8.41$\%$ and 8.40$\%$ for $CdGa_2X_4$ (X = S, Se, Te) respectively. The sphere radii and the positions are chosen in such a way that space filling in the unit cell is achieved without exceeding a sphere overlap of 16$\%$. The atomic and interstitial sphere radii are calculated and given in table 1, table 2 and table 3 respectively for $CdGa_2X_4$. Total nine empty spheres of four different radii (E, E1, E2 and E3) are introduced for proper overlap of the spheres in all three systems. We obtain the ground state equilibrium values of lattice parameters, anion positions (u) and bond lengths by structural relaxation. It is done by minimizing  total energy using the above first principle procedure. We calculate the Bulk modulus by fiiting energies versus cell volume data to Murnaghan equation of state.
\begin{table}
\begin{center}
\caption{ Radii and positions of atomic and empty spheres in $CdGa_2S_4$.}
\begin{tabular}{lccccc}
\hline
atoms & x & y & z & sphere radii (a.u.) \\ 
\hline
Cd & 0.000 & 0.000 & 0.000 & 2.828 \\
 E & 0.000 & 0.500 & -0.497 & 2.332 \\
Ga1& 0.000 & 0.000 & 0.995 & 2.585 \\
Ga2& 0.500 & 0.000 & -0.497& 2.642 \\ 
S  & 0.248 & -0.270 & -0.265 & 2.572 \\ 
S & -0.248 & 0.270 & -0.265 & 2.572\\
S & 0.230 & 0.252 & -0.730 & 2.572\\
S & -0.230 & -0.252 & 1.259 & 2.572 \\
E1& 0.500 & 0.000 & -0.041& 2.582 \\ 
E1& 0.500 & 0.000 &-0.953& 2.582\\ 
E2& 0.270 & -0.248 & 0.265 & 2.572\\
E2& -0.270 & 0.248 & 0.265 & 2.572 \\
E2& -0.252 & -0.230 & 0.730 & 2.572\\
E2 &0.252 & 0.230 & -1.259 & 2.572 \\ 
E3 & 0.000 & 0.000 & 0.456& 2.367\\ 
E3 & 0.000 & 0.000 &-0.456& 2.367 \\ 
\hline
\end{tabular}\\
\end{center}
\end{table}

\begin{table}
\begin{center}
\caption{Radii and positions of atomic and empty spheres in $CdGa_2Se_4$.}
\begin{tabular}{lccccc}
\hline
atoms & x & y & z & sphere radii (a.u.) \\ 
\hline
Cd& 0.000 & 0.000 & 0.000 & 2.827 \\ 
E & 0.000 & 0.500 & -0.499& 2.400\\ 
Ga1& 0.000 & 0.000 & 0.997 & 2.646\\
Ga2& 0.500 & 0.000& -0.499& 2.618 \\ 
Se &  0.249 &-0.264 &-0.265& 2.727\\ 
Se & -0.249 & 0.264 &-0.265&2.727 \\ 
Se & 0.236 & 0.251 &-0.732& 2.727 \\ 
Se & -0.236 & -0.251 & 1.263& 2.727\\ 
E1&  0.264 &-0.249 & 0.265& 2.726\\ 
E1& -0.264 & 0.249 & 0.265& 2.726\\ 
E1 &-0.251 &-0.236 & 0.732&2.726 \\
E1&  0.251 & 0.236 &-1.263& 2.726\\
E2 & 0.500 & 0.000 &-0.028 & 2.647\\
E2 & 0.500 & 0.000 &-0.970 &2.647\\
E3 & 0.000 & 0.000 & 0.471& 2.427\\
E3 & 0.000 & 0.000 &-0.471&2.427\\
\hline
\end{tabular}\\
\end{center}
\end{table}

\begin{table}
\begin{center}
\caption{Radii and positions of atomic and empty spheres in $CdGa_2Te_4$.}
\begin{tabular}{lccccc}
\hline
atoms & x & y & z & sphere radii (a.u.) \\ 
\hline
Cd & 0.000 & 0.000 & 0.000 & 2.895\\
E  & 0.000 & 0.500 &-0.494 & 2.589\\
Ga1& 0.000 & 0.000 & 0.989 & 2.666\\
Ga2& 0.500 & 0.000 &-0.494 & 2.791\\
Te & 0.251 &-0.264 &-0.253 & 3.016\\
Te &-0.251 & 0.264 &-0.253 & 3.016\\
Te & 0.236 & 0.249 &-0.736 & 3.016\\
Te &-0.236 &-0.249 & 1.242 & 3.016\\
E1 & 0.264 &-0.251 & 0.253 & 3.015\\
E1 &-0.264 & 0.251 & 0.253 & 3.015\\
E1 &-0.249 &-0.236 & 0.736 & 3.015\\
E1 & 0.249 & 0.236 &-1.242 & 3.015\\
E2 & 0.500 & 0.000 & 0.000 & 2.681\\
E2 & 0.000 & 0.500 & 0.000 & 2.681\\
E3 & 0.000 & 0.000 & 0.467 & 2.613\\
E3 & 0.000 & 0.000 &-0.467 & 2.613\\
\hline
\end{tabular}\\
\end{center}
\end{table}

\subsection{Optical Properties}
Optical properties can be determined by the dielectric function $\varepsilon(\omega)$ which is a complex quantity. $\varepsilon(\omega) = \varepsilon_1 (\omega) + i\varepsilon_2 (\omega)$ where $\varepsilon_1 (\omega)$ $\&$ $\varepsilon_2 (\omega)$ are the real and imaginary parts of the dielectric function. There are two contribution to $\varepsilon(\omega)$. One is due to intraband transitions and another is due to interband transitions. The contribution from intraband transitions is important for metals and semimetals. But interband transitions are important in case of semiconductors. To calculate the direct interband contributions to the imaginary part of the dielectric function $\varepsilon_2 (\omega)$, all the possible transitions from the occupied to the unoccupied states within the irreducible Brillouin zone must be summed up. The interband transitions can be of two types i.e. direct and indirect transitions. Here we neglect the indirect interband transitions which involve electron phonon scattering and are expected to give only a small contribution to $\varepsilon(\omega)$.
The expression for the imaginary part of the dielectric function is given by the well known Kubo formula for the direct interband transitions \cite{33,34}
\begin{equation}
\epsilon_{2}^\gamma(\omega)=\eq \frac{-8{\pi}^{2} e^{2}}{ 2m^{*^{2}} \Omega} \frac{1}{{\omega}^2}
\sum_{i}\sum_{f} {\vert \langle \psi_{f} \vert {\mathaccent 94 e}
_{\gamma}\cdot{\mathaccent 94 p} \vert \psi_{i}\rangle \vert}^{2}F_{i}(1-F_{f})
\delta (E_{f}-E_{i}-\hbar \omega )
\end{equation}
 where,
$ m^{*}$ is the effective mass of the electron,$ \Omega $ is the volume of the sample, i and f refer to the initial and final states respectively, $ \gamma$ refers to the direction of polarization of the incoming photon, ${\mathaccent 94 p}$ is the momentum of the electron and $ F_{i,f}$ is the occupation probability of the initial and final states respectively. For semiconductors, i lies in the valence band and f in the conduction band and at $0K$ we have $F_{i} = 1$ and $F_{f} = 0$. We can obtain the real part of the dielectric function $\epsilon_1(\omega)$ from Kramers-Kronig relation \cite{33}.
\begin{equation}
\epsilon_{1}(\omega) = \eq 1 + \pls \frac{2}{\pi}\int_{0}^{\infty} \frac{(\omega^{\prime}\mns\omega)
 \epsilon_{2}(\omega^{\prime})}{\omega^{\prime^2}\mns\omega{^2}}d\omega^{\prime}
\end{equation}
A number of methods have been proposed for calculating optical properties within the frame work of the LMTO \cite{35,36,37,38}  for both metals \cite{35,36} and semiconductors \cite{37}. In this  study we use the method developed by Hobbs et.al. \cite{38} which avoids the determination of gradient operator. This method allows for the inclusion of non-local potentials also in the Hamiltonian. The momentum matrices which appear in the Kubo formula, are written in terms of Gaunt coefficient and potential parameters which are defined within the TB-LMTO method. We calculate these Gaunt coefficient separately. \\
The imaginary part of dielectric function, optical matrix elements and the joint density of states can be calculated by performing an integration over the Brillouin zone (k-space integration). For all these calculation, the tetrahedron method is used within TB-LMTO formalism. In this method, the dielectric function is expressed as an integral over the constant-energy surface , $E_c(k) - E_v(k)$. The eigen values and eigen vectors are then calculated on a mesh in the irreducible Brillouin zone. This zone is devided in to tetrahedra of equal volume and the mesh of k-points (maximum 512 points) defines the corners of each tetrahedron. The interpolated function is continious at the boundaries of the tetrahedra. The irreducible Brillouin zone is completely divided into tetrahedra. To calculate different necessary formulae for optical properties, the output data from the TB-LMTO band structure calculations are used.\\
\section{Result and discussion}
\subsection{Structural properties :}
One unit cell of $CdGa_2X_4$ chalcopyrite is shown in figure 1. It is basicaaly a supercell of two zinc-bende unit cells placed one on top of the other with a $90^0$ rotation about c-axis. There are two Cd, two vacancies, four Ga and eight X atoms per unit cell. The positions of various atoms in the tetragonal unit cell are (in Wyckoff notation): Cd on 2a site (0 0 0), Ga1 on 2b site (0 0 0.5), Ga2 on 2d site (0 0.5 0.75), vacancy on 2c site (0 0.5 0.25) and X on 8g site ($u_x\  u_y\  u_z$). Here ‘$u_x$’, ‘$u_y$’ and ‘$u_z$’ are anion displacement parameters along three axes. The space group of the defect chalcopyrite is $I{\bar 4}(S_4^2)$. Unit cell with $\eta (= c/2a) = 1$, $u_x$, $u_y$ and $u_z$ equal to 0.25, 0.25 and 0.125 respectively, is referred as ideal structure (ideal case) \cite{20,21,22,23,24}. Anion shifts along all the three directions in these systems, unlike only along x-direction found in the case of pure chalcopyrites \cite{20,22}. This new position of anion is called anion displacement. This is due to the reduction in symmetry in these systems. Therefore, all cations-anion bond lengths are inequivalent in these systems. Calculated bond-legnths tabulated in the table 4 show  anion acquires an equilibrium position closer to the vacancy than to the other cations as expected.\\
\begin{figure}
\centering
\hspace*{-1.5cm}
\includegraphics[scale=0.55]{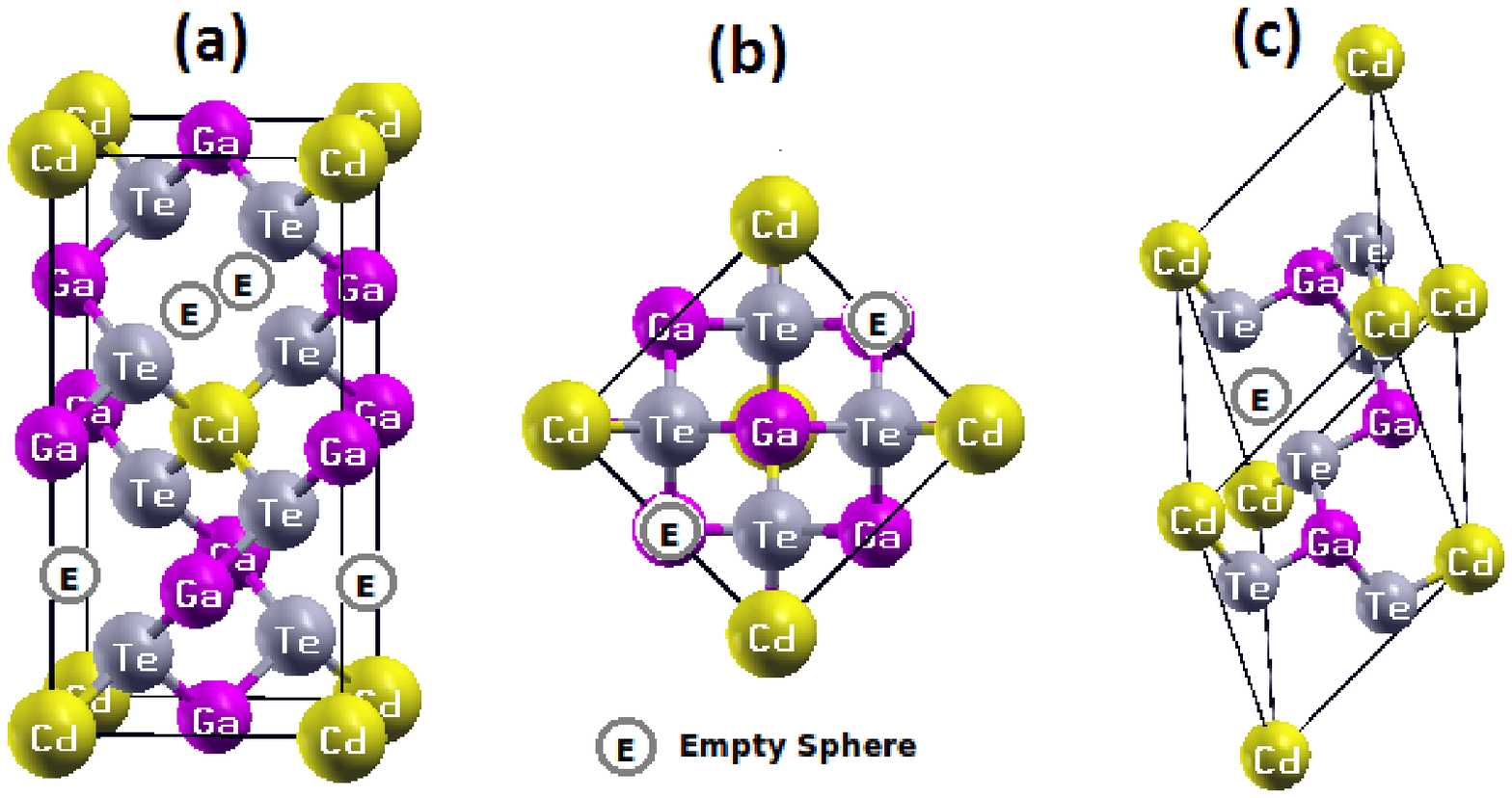}
\caption{One unit cell of $CdGa_2X_4$ chalcopyrie semiconductor (a) front view (b) top view. (c) primitive cell of $CdGa_2Te_4$.}
\end{figure}
Table 5 shows our result of lattice parameters ‘$a$’ \& ‘$c$’, tetragonal distortion, and anion displacement agree quite well with experimental and other calculated results. Table 6 shows our result of bulk moduli agree well with the other theoretical and experimental results also.
\begin{table}
\begin{center}
\caption{Bond lengths(R) in $\AA{}$}
\begin{tabular}{lccc}
\hline
$CdGa_2S_{4}$ &Present Work & Experimental$^a$& Other Theoretical$^b$ \\
\hline
$R_{Cd-S}$ & 2.48 & 2.52 & 2.57 \\
$R_{Ga1-S}$& 2.37 & 2.29 & 2.31 \\
$R_{Ga2-S}$& 2.40 & 2.33 & 2.31\\
$R_{Vacancy-S}$& 2.26 & - & - \\
\hline
$CdGa_2Se_4$&Present Work & Experimental$^c$& Other Theoretical $^b$\\
\hline
$R_{Cd-Se}$ & 2.57 & 2.63 & 2.70 \\
$R_{Ga1-Se}$& 2.48 & 2.41 & 2.41  \\
$R_{Ga2-Se}$& 2.47 & 2.41 & 2.41 \\
$R_{Vacancy-Se}$& 2.37 & - & - \\
\hline
$CdGa_2Te_4$&Present Work & Experimental$^d$& Other Theoretical$^e$ \\
\hline
$R_{Cd-X}$ & 2.74 & - & 2.88\\
$R_{Ga1-X}$& 2.63 & - & 2.68\\
$R_{Ga2-X}$& 2.69 & - & 2.66\\
$R_{Vacancy-X}$& 2.59 & - & - \\
\hline
\end{tabular}
\end{center}
$^a$ Ref.[10]; $^b$ Ref.[14]; $^c$ Ref.[39]; $^d$ Ref.[40]; $^e$ Ref.[19]
\end{table}

\begin{table}
\begin{center}
\caption{Structural parameters.}
\begin{tabular}{ccccccc}
\hline
Compounds& &a& c&$u_x$&$u_y$&$u_z$\\
 & &$(\AA)$&$(\AA)$ & & & \\
\hline
$CdGa_2S_4$&Present work&5.51&10.94&0.270&0.248&0.133\\
           &Experiment$^a$&5.53&10.16&0.271&0.261&0.140\\
           &Other theory$^b$&5.64&10.34&0.268&0.271&0.136 \\
\hline
$CdGa_2Se_4$& Present work&5.72&11.41&0.264&0.249&0.133 \\
            &Experiment$^c$&5.73&10.70&0.275&259&139\\    
          &Other theory$^b$&5.88&10.78&0.273&0.270&0.138   \\
\hline
$CdGa_2Te_4$&Present work & 6.18& 12.22&0.264 &0.251& 0.128\\
            &Experiment$^d$ &6.10 &11.70&0.270& 0.260&0.135 \\
            &Other theory$^e$& 6.27&12.00&0.272&.255&137\\
\hline
\end{tabular}
\end{center}
$^a$ Ref.[10]; $^b$ Ref.[14]; $^c$ Ref.[39]; $^d$ Ref.[40]; $^e$ Ref.[19]
\end{table}

\begin{table}
\begin{center}
\caption{Bulk modulus B in GPa.}
\begin{tabular}{lcccc}
\hline
Systems & Present work & Experimental & Other theoretical \\ 
\hline
$CdGa_2S_{4}$ & 57.10 & 64.0 $^a$ & 58.44$^b$\\
$CdGa_2Se_4$ & 38.32  & 41.0$^c$ & 36.13$^b$\\
$CdGa_2Te_4$ &31.00  & - & -\\
\hline
\end{tabular}
\end{center}
$^a$ Ref.[10]; $^b$ Ref.[14]; $^c$ Ref.[8]
\end{table}

\subsection{Structural effect on electronic properties}
Band structure, total density of states (TDOS) and partial density of states (PDOS) are calculated to study the detailed electronic properties of $CdGa_2X_4$. The Fermi level is set at zero energy in the band structure and total DOS plots and marked by dashed horizontal line.  E, E1, E2 and E3 represent empty spheres of different radii in band structure diagram which are introduced at appropriately chosen interstitial sites. Band strucrure diagram of $CdGa_2S_4$ for non-ideal structure is only shown here in the figure 2 for reference. A similar picture arise in the case of other two systems. The partial density of states shown in the figure 2 show that p-d hybridization play important role in band formation.\\
 
The valence band maximum and conduction band minimum are located at the center of brillouin zone denoted as ``G'' ($\Gamma$ point). This indicates that these are direct band gap semiconductors. Table 7 shows our result of band gaps underestimate the experimental result and are closed to some of the other theoretical results. This die to LDA. It is known that LDA underestimates band gap by 30-50$\%$ \cite{41}. There are are better functionals avialable like hybrid \cite{31} for accurate calculation of band gap. But our motivation here is not to calculate accuarate band gap. Our main motivation is to show effects of structural distortion on electronic and optical properties. The choice of LDA does not affect such a study.
\begin{figure}
\begin{center}
\begin{tabular}{cc}
{\rotatebox{0}{\resizebox{8.0cm}{8.0cm}{\includegraphics{{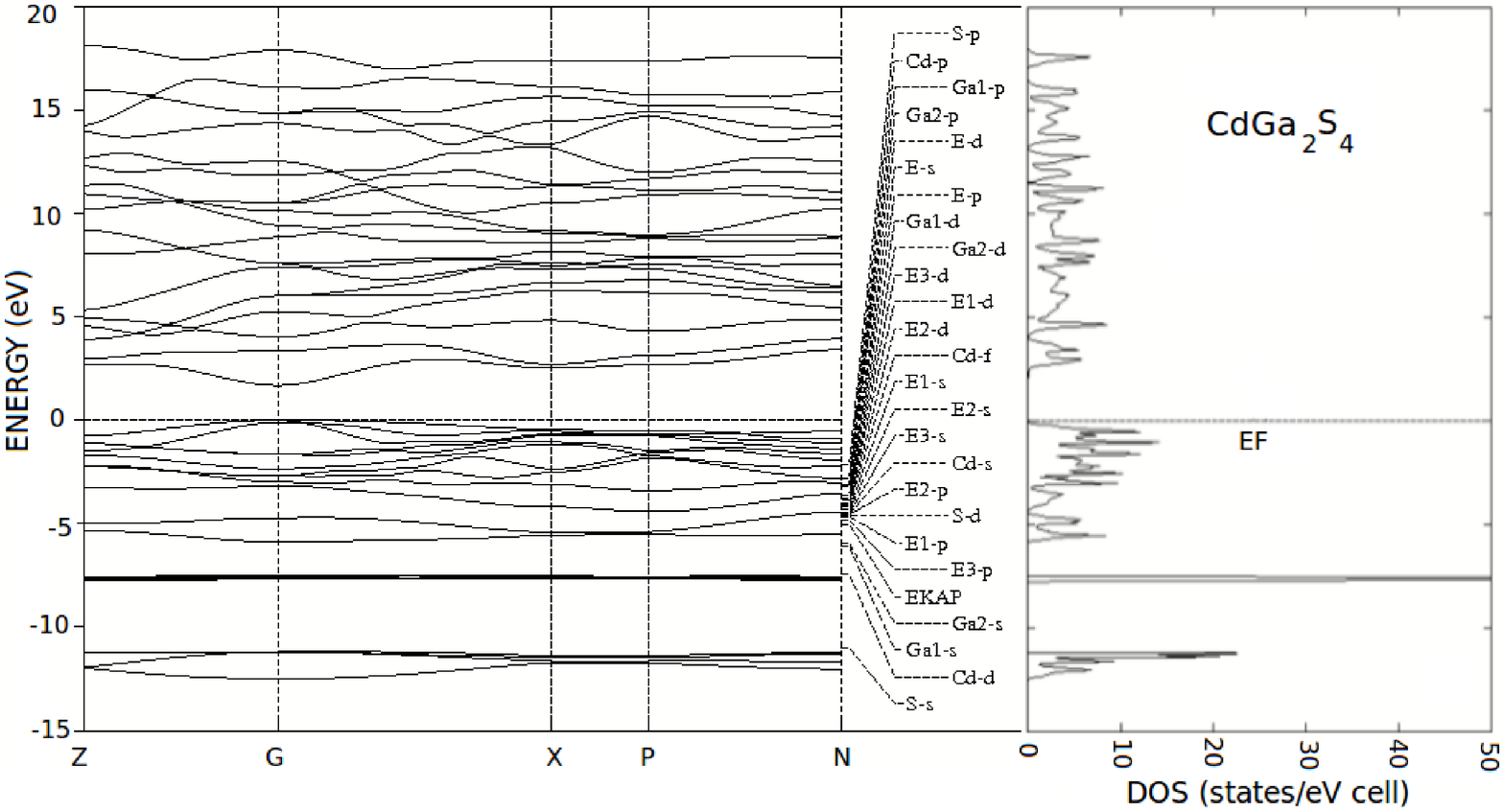}}}}} &
{\rotatebox{0}{\resizebox{6.0cm}{8.0cm}{\includegraphics{{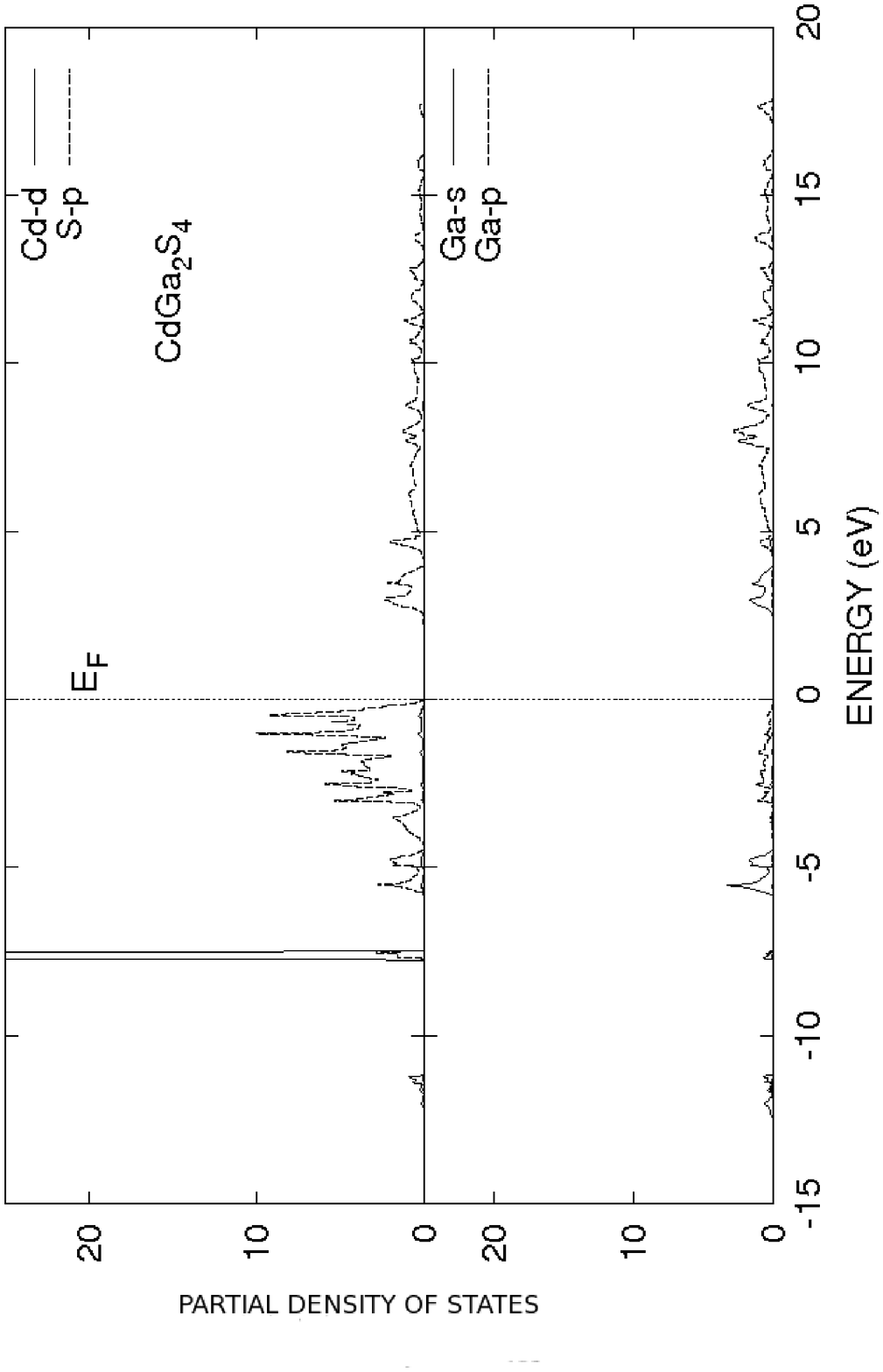}}}}} \\
{\rotatebox{-90}{\resizebox{7.0cm}{7.0cm}{\includegraphics{{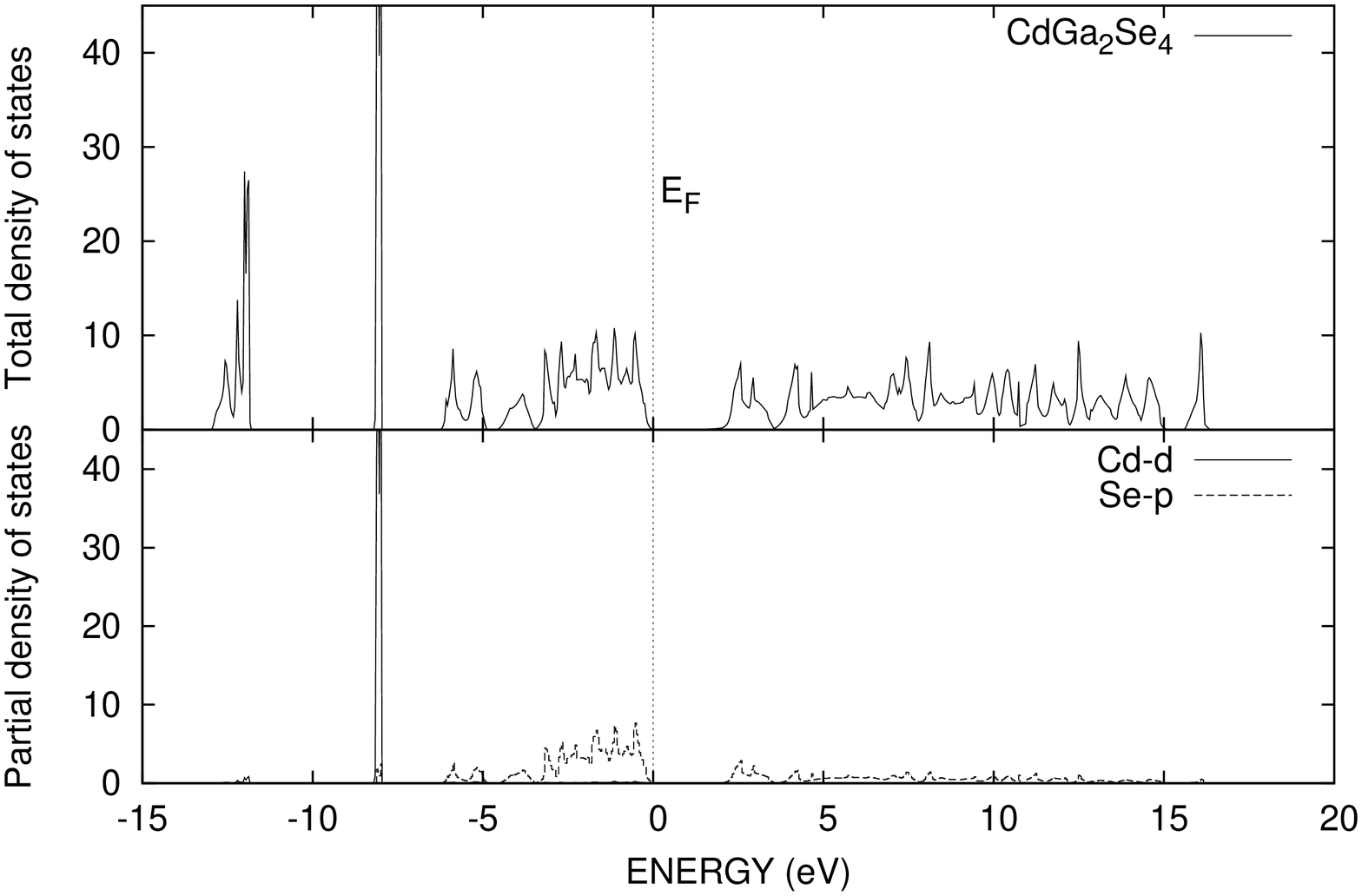}}}}} &
{\rotatebox{-90}{\resizebox{7.0cm}{7.0cm}{\includegraphics{{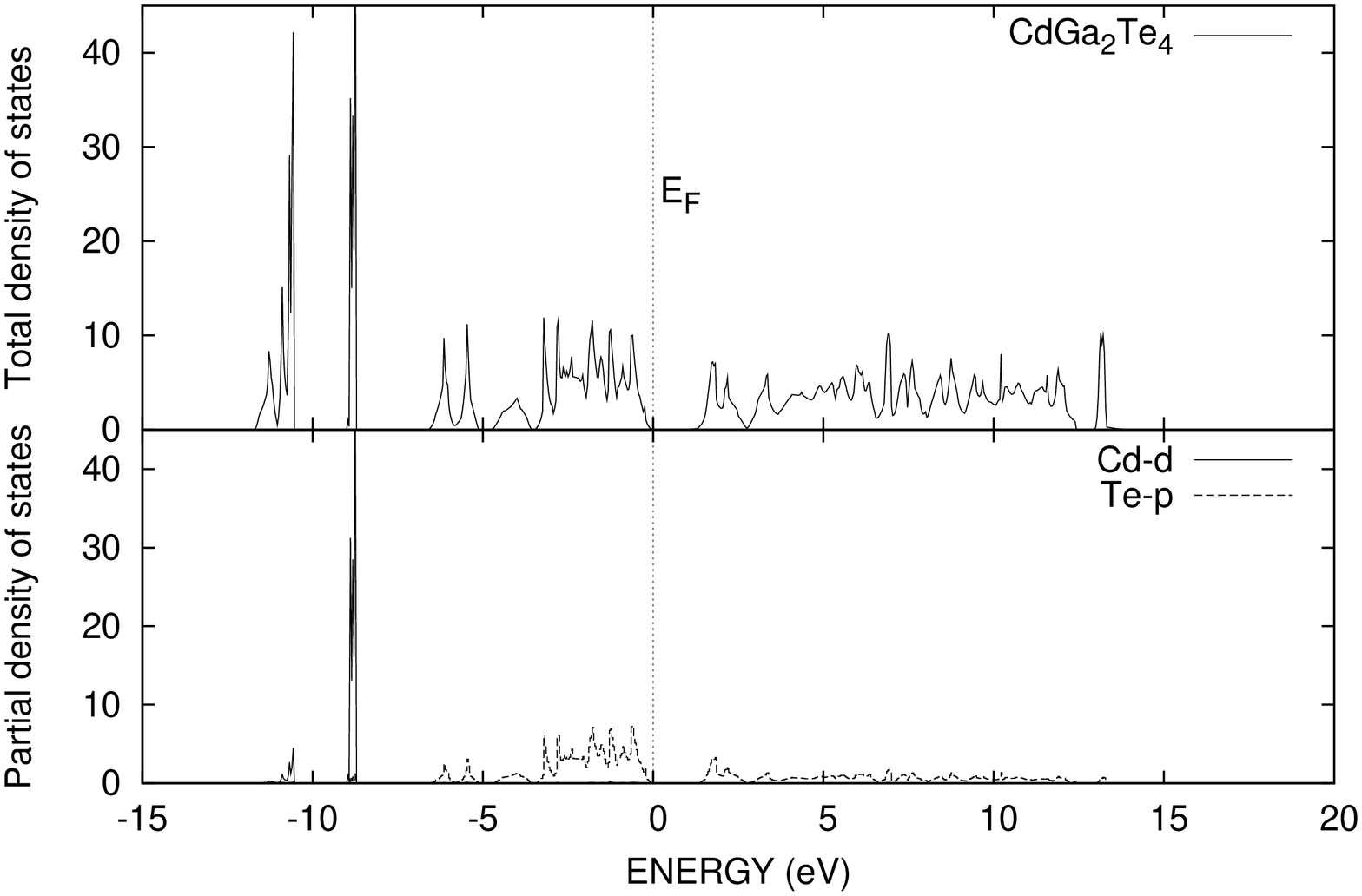}}}}}
\end{tabular}
\caption{(Top panel): Band structure and density of states for non-ideal $CdGa_2S_4$.. (Bottom panel): Density of states for non-ideal $CdGa_2Se_4$ \& $CdGa_2Te_4$}
\end{center} 
\end{figure}

\begin{figure}
   \begin{center}
{\rotatebox{-90}{\resizebox{9.0cm}{12.0cm}{\includegraphics{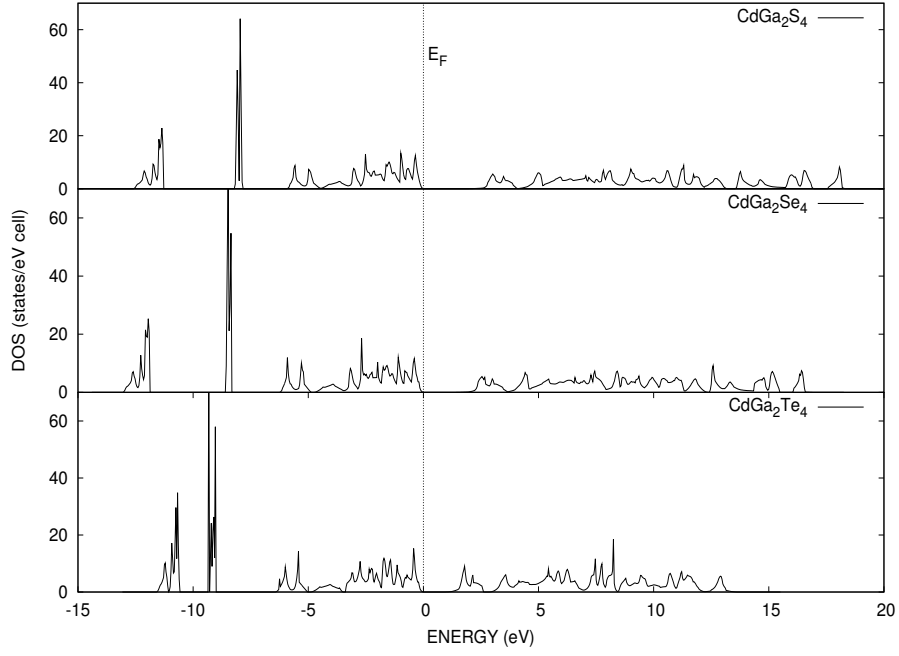}}}}
\caption{Total DOS for ideal $CdGa_2X_4$ ($X=S,Se,Te$)}
   \end{center} 
\end{figure}

We had earlier carried out quantitative study of effect of structural distortion on band gap in a series of chalcopyrite semiconducting compounds \cite{22,23,24}. We apply similar method here  in the present study also. Table 8 shows $\%$ of increment in band gap due to the structural distortion. Figures 2 \& 3 show no significant differences in the structures of TDOS in ideal and non-ideal cases.  A very little differences are visible in the conduction bands in the cases of $CdGa_2S_4$ \& $CdGa_2Se_4$. The conduction band width increases in case of ideal $CdGa_2S_4$ compared to the non-ideal case. Whereas significant differences are visible in both valence band and conduction bands for ideal and non-ideal $CdGa_2Te_4$. We also observe that the effect of structural distortion on band gap increases from $CdGa_2S_4$ to $CdGa_2Te_4$. 

\begin{table}
\begin{center}
\caption{Energy band gap $E_g$ (eV).}
 \begin{tabular}{@{}lcccc}
\hline
 Systems &  $E_g$  & $E_g{^{exp}}^a$  & $E_g^{other}$\\
         &(eV) & (eV) & (eV\\
\hline
$CdGa_2S_4$  & 2.20 & 3.25-3.44 & 3.30$^b$, 2.11$^c$     \\
$CdGa_2Se_4$ & 1.75 & 2.57& 2.76$^b$, 1.46$^c$  \\
$CdGa_2Te_4$ &1.25  & 1.50 & 1.90$^b$\\
\hline
\end{tabular}\\
$E_g^{exp}$ : experimental, $E_g^{other}$ : other calculated gap.\\
$^{a}$ Ref.\cite{42}, $^{b}$ Ref.\cite{9}; $^{c}$ Ref.\cite{14}; 
\end{center}
\end{table} 
\begin{table}
\centering
\caption{Effect of structural distortion on band gap (eV).}
\begin{tabular}{@{}lccc}
\hline
Systems &  Ideal  &  Non-ideal & Increment in  \\ 
        &   (eV)  &  (eV)      & band gap ($\%$) \\ \hline
$CdGa_2S_4$ &2.12 & 2.20 & 3.63 \\
$CdGa_2Se_4$& 1.68 &1.75& 4.0\\
$CdGa_2Te_4$& 1.14 &1.25&8.8\\
\hline
\end{tabular}
\end{table}

\subsection{Structural effects on optical properties}

There are two important factors in the Kubo formula for imaginary part of dielectric function $\epsilon_2 (\omega)$. One is joint density of states (JDOS) and the other is the ``square of the optical matrix elements'' (OME) (transition probabilities or electron photon interaction matrix elements). The various peaks in $\epsilon_2 (\omega)$ come from the joint density of states for values of $E_c$ \& $E_v$ at which slopes of valence band and conduction band are equal (Von Hove singularities). This occurs at high symmetry points in the Brillounin zone. For other values where slope of $E_c$ is not equal to slope of  $E_v$, transitions may occur at any general points in Brillouin zones. But OME is a difficult factor to be interpreted accurately. In almost all the works/literature, there is no explicit calculation of this quantity to see its effect on optical absorption of chalcopyrite semiconductors. The observed or calculated absorption co-efficient/imaginary part of the dielectric function are generally interpreted as transition from various critical points. Therefore majority of the work correlates $\epsilon_2$ with joint density of states.\\

Chalcopyrite semiconductors have tetragonal structure and therefore its optical properities show anisotropic nature. Therefore oprical response functions are different when incoming photon is polarised along c-axis than when it is polarised along perpendicular to c-axis.\\  

Figure 4 shows no significant differences in JDOS between ideal and non-ideal structures of $CdGa_2S_4$.  The magnitude of JDOS is almost equal in both the structures . Similar results are also observed in other two systems. But figure 5 shows remarkable differences in OME between the two structures. The magnitude of optical matrix elements is almost 1.75 times higher in ideal case than the non-ideal. Lot of structures which are present in the ideal structures get drastically reduced in non-ideal structure. Therefore any differences in $\epsilon_2$  between the two structures are mainly coming from changes in OME. For example we notice from figure 6 that the magnitude of the imaginary part of dielectric function ($\epsilon_2$) in the case of ideal $CdGa_2S_4$ is almost double compared to non-ideal $CdGa_2S_4$. The highest peak in OME and $\epsilon_2$ is found at the same energy. In case of ideal $CdGa_2Te_4$ the magnitude of OME increases up to 7 and the peak found at energy 2.5 eV in non-ideal case splits in to two sharp peaks. So  structures in OME are clearly reflected in $\epsilon_2$ for both $CdGa_2Se_4$ and $CdGa_2Te_4$. Significant differences are also observed in the case of real part of dielectric function as can be seen in the figure 7. Figures 5, 6 \& 7 show also anisotropic nature of the optical properties and they are more prominent for the non-ideal structure than the ideal.\\
We also find that static dielectric constants and refractive indices increase for ideal structure of $CdGa_2X_4$. Table 9 \& 10 show the static dielectric constant $\epsilon_1(0)$ and refractive index $n(0)$ for non-ideal and  ideal $CdGa_2X_4$ respectively. Our calculated $\epsilon_1$ and $\epsilon_2$ for $CdGa_2Te_4$ agree well with the result of Ozaki et.al. \cite{45}
\begin{figure}
\centering
\begin{tabular}{cc}
{\rotatebox{-90}{\resizebox{7.0cm}{7.0cm}{\includegraphics{{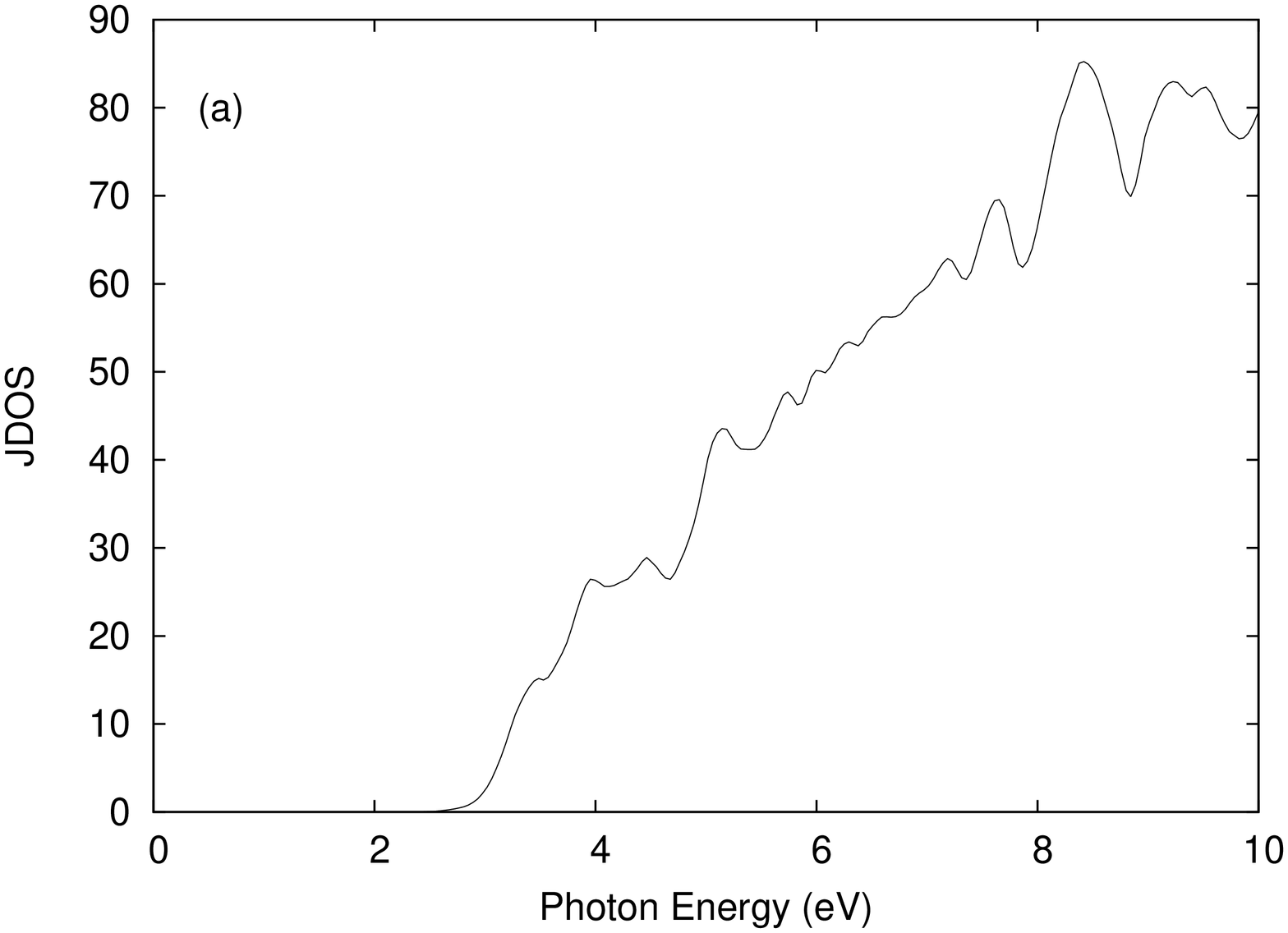}}}}} & {\rotatebox{-90}{\resizebox{7.0cm}{7.0cm}{\includegraphics{{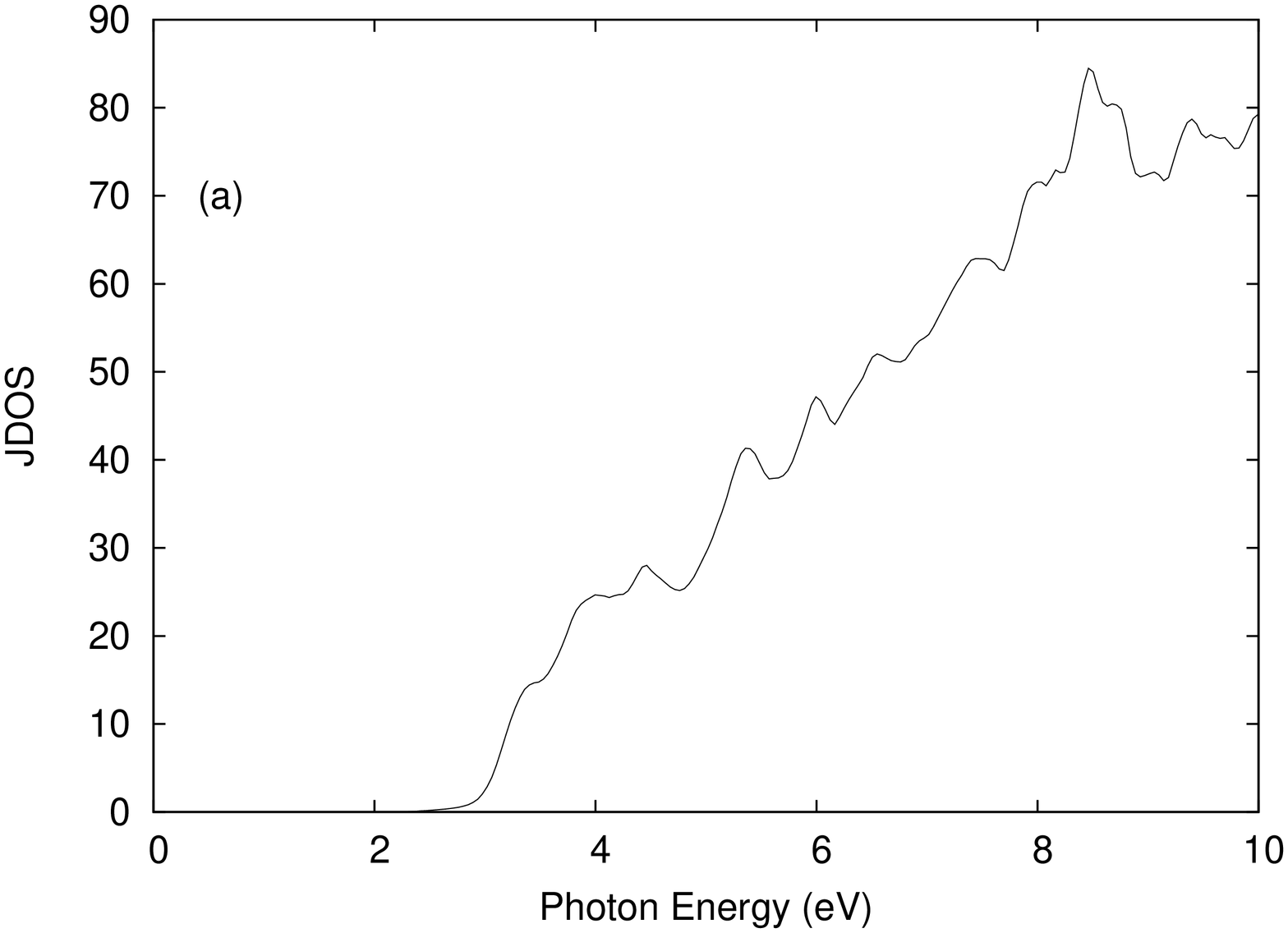}}}}} 
\end{tabular}
\caption{(Left panel) : Joint density of staes for non-ideal; (right panel): for ideal  $CdGa_2S_4$}
\end{figure}

\begin{figure}
\centering
\begin{tabular}{cc}
{\rotatebox{-90}{\resizebox{7.0cm}{7.0cm}{\includegraphics{{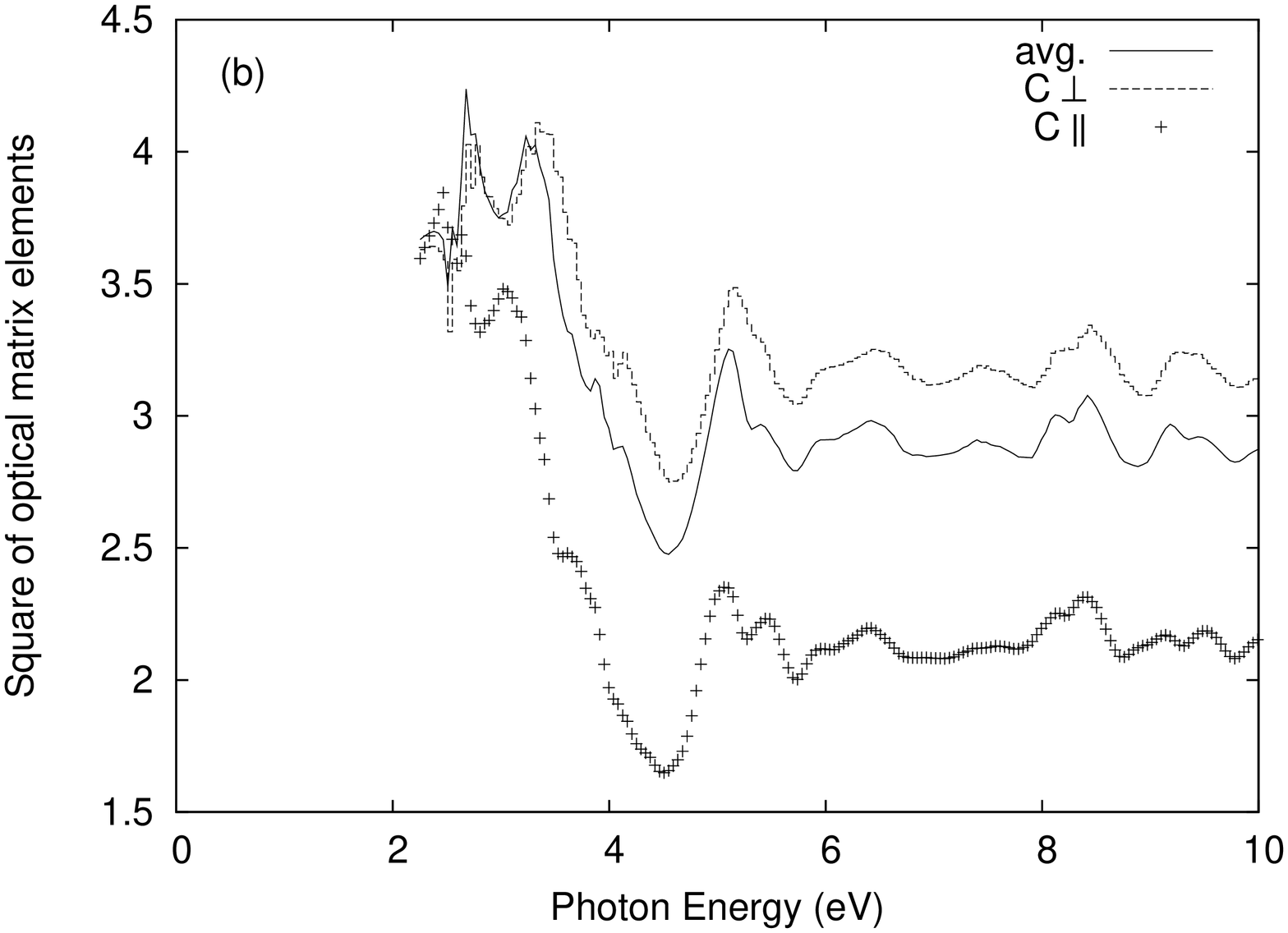}}}}} & {\rotatebox{-90}{\resizebox{7.0cm}{7.00cm}{\includegraphics{{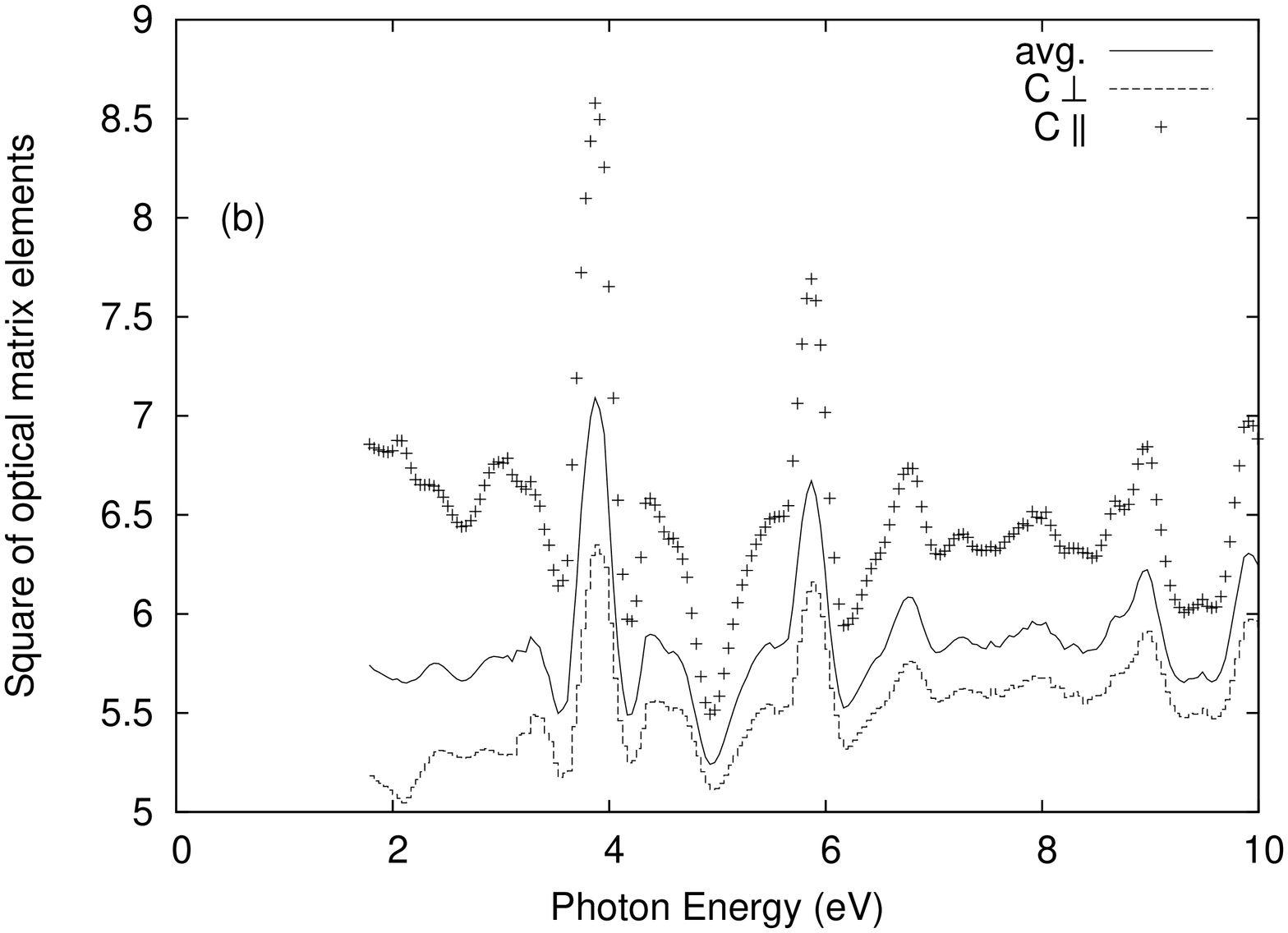}}}}}\\
{\rotatebox{-90}{\resizebox{7.0cm}{7.0cm}{\includegraphics{{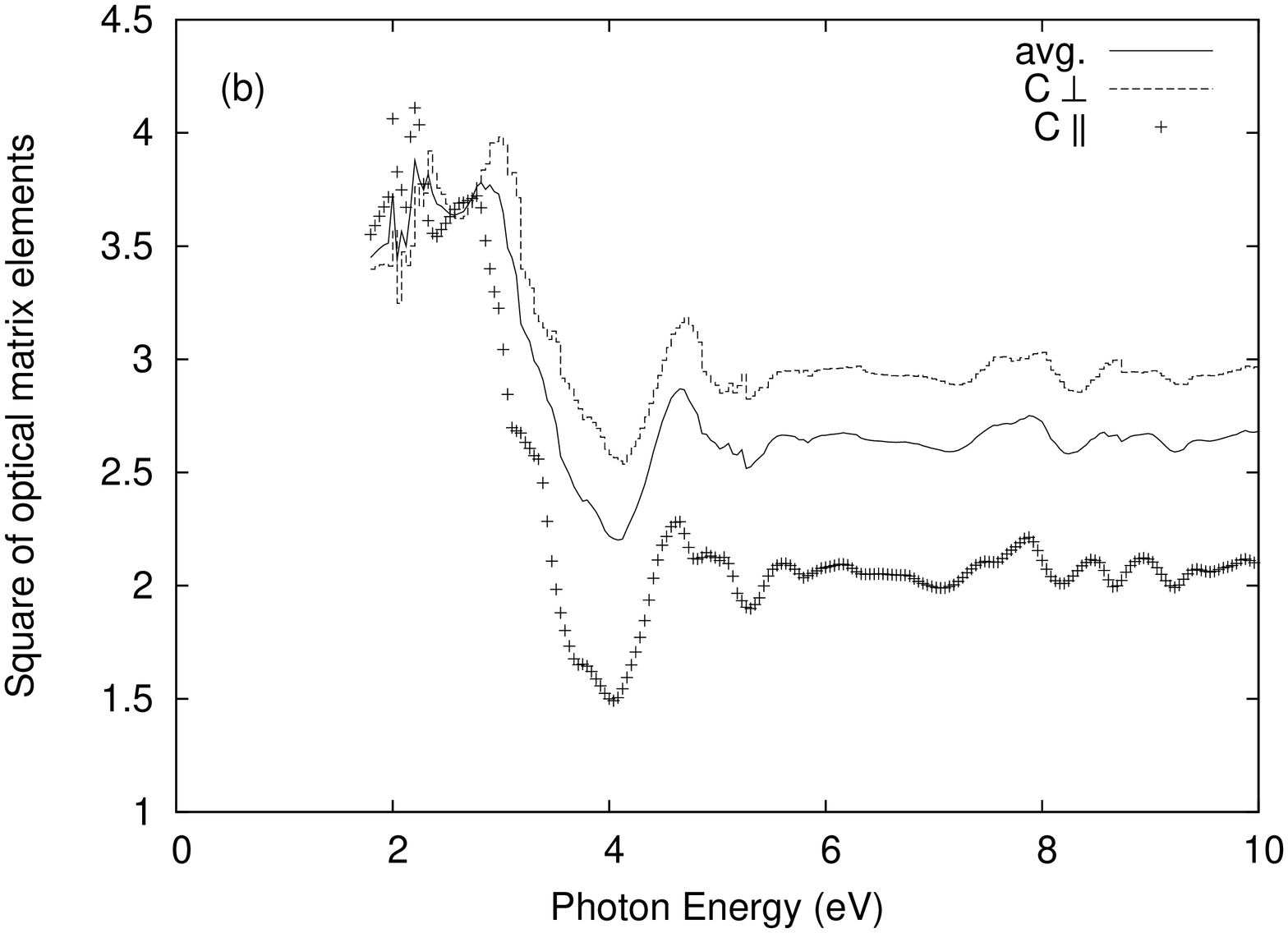}}}}} & {\rotatebox{-90}{\resizebox{7.0cm}{7.0cm}{\includegraphics{{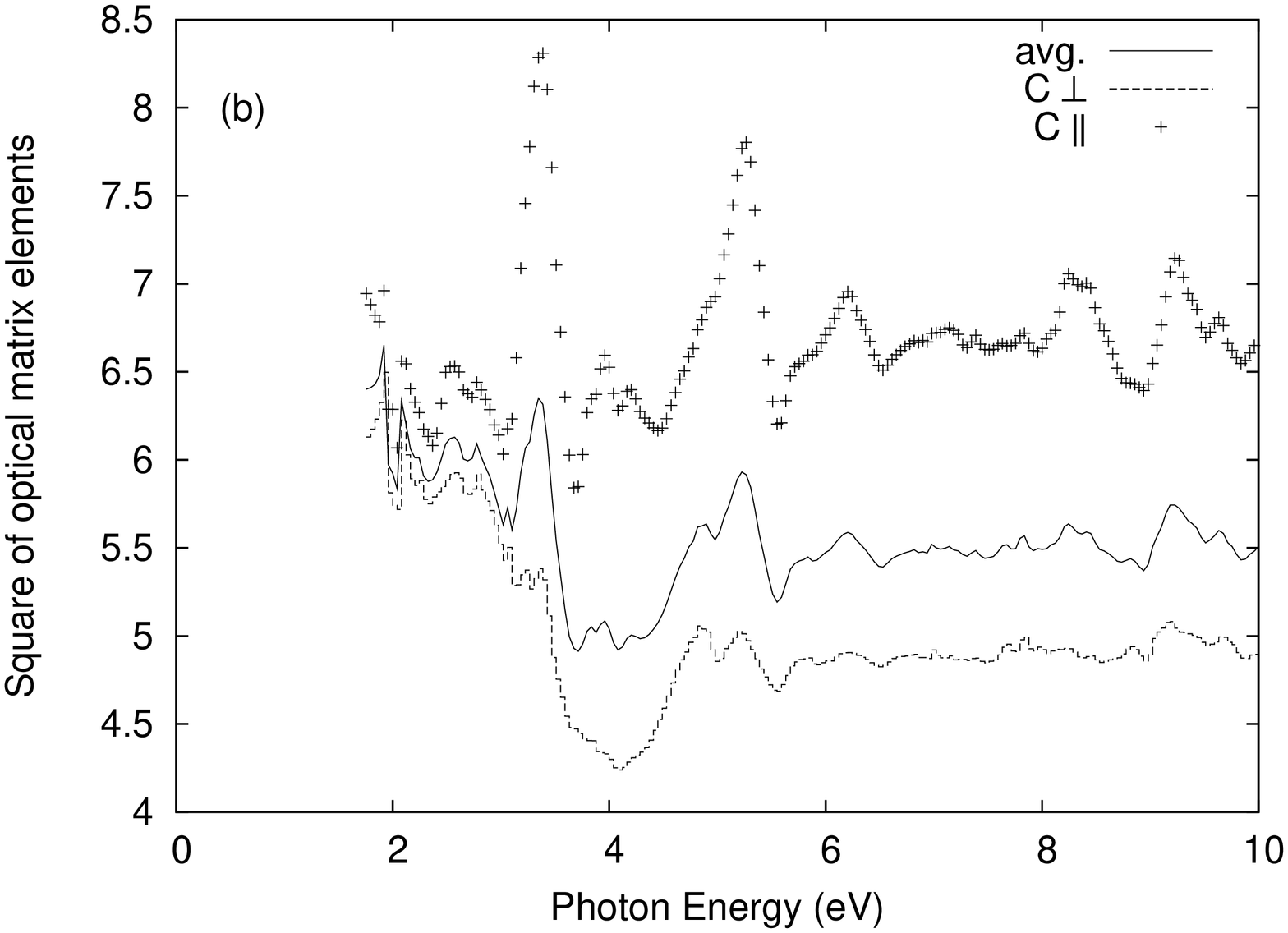}}}}}\\
{\rotatebox{-90}{\resizebox{7.0cm}{7.0cm}{\includegraphics{{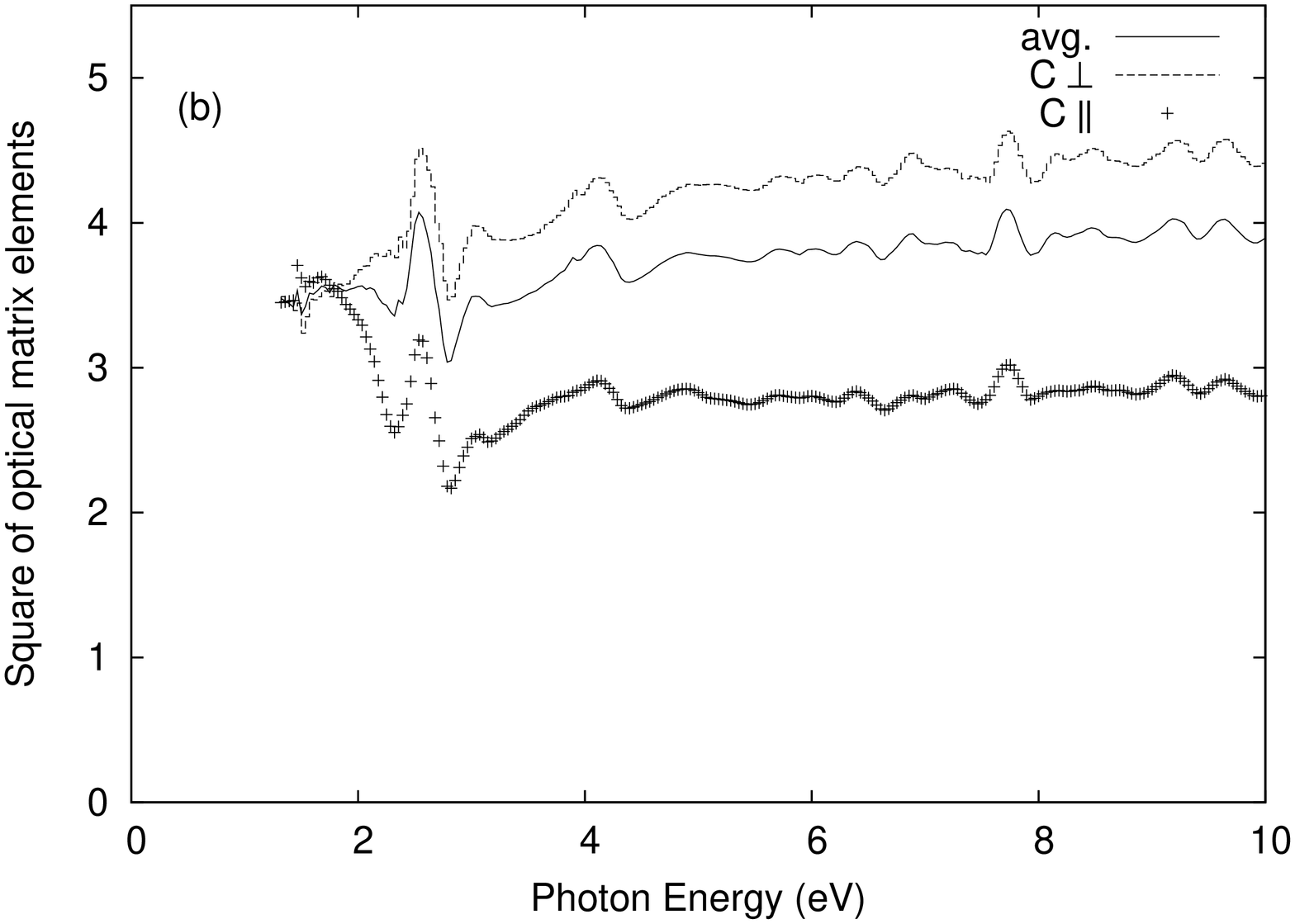}}}}} & {\rotatebox{-90}{\resizebox{7.0cm}{7.0cm}{\includegraphics{{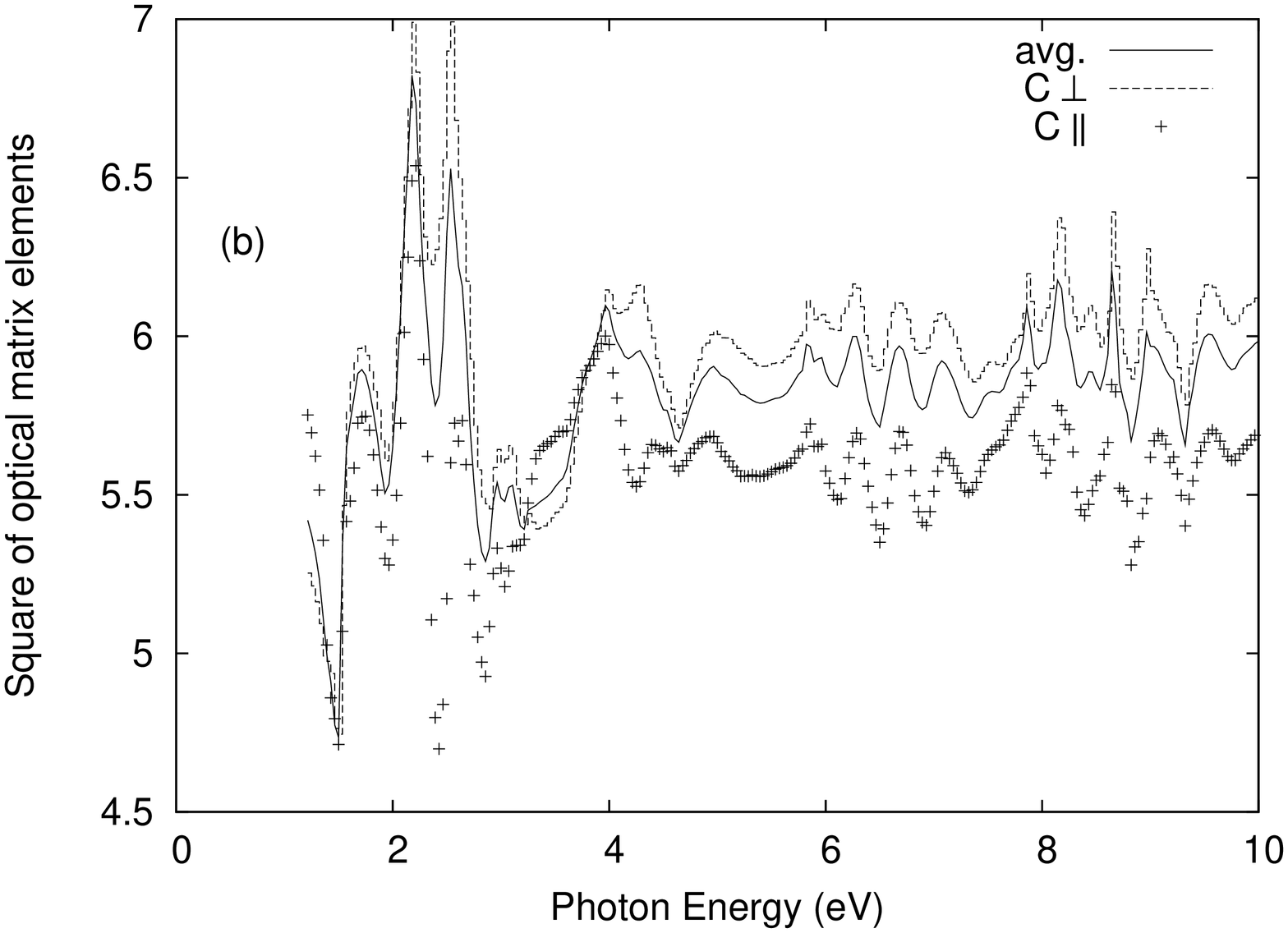}}}}}
\end{tabular}
\caption{(Left panel) : Optical matrix elements for non-ideal (right panel) : for ideal $CdGa_2X_4$ ; $X=S,Se,Te$ respectively}
\end{figure}

\begin{figure}
\centering
\begin{tabular}{cc}
{\rotatebox{-90}{\resizebox{7.0cm}{7.0cm}{\includegraphics{{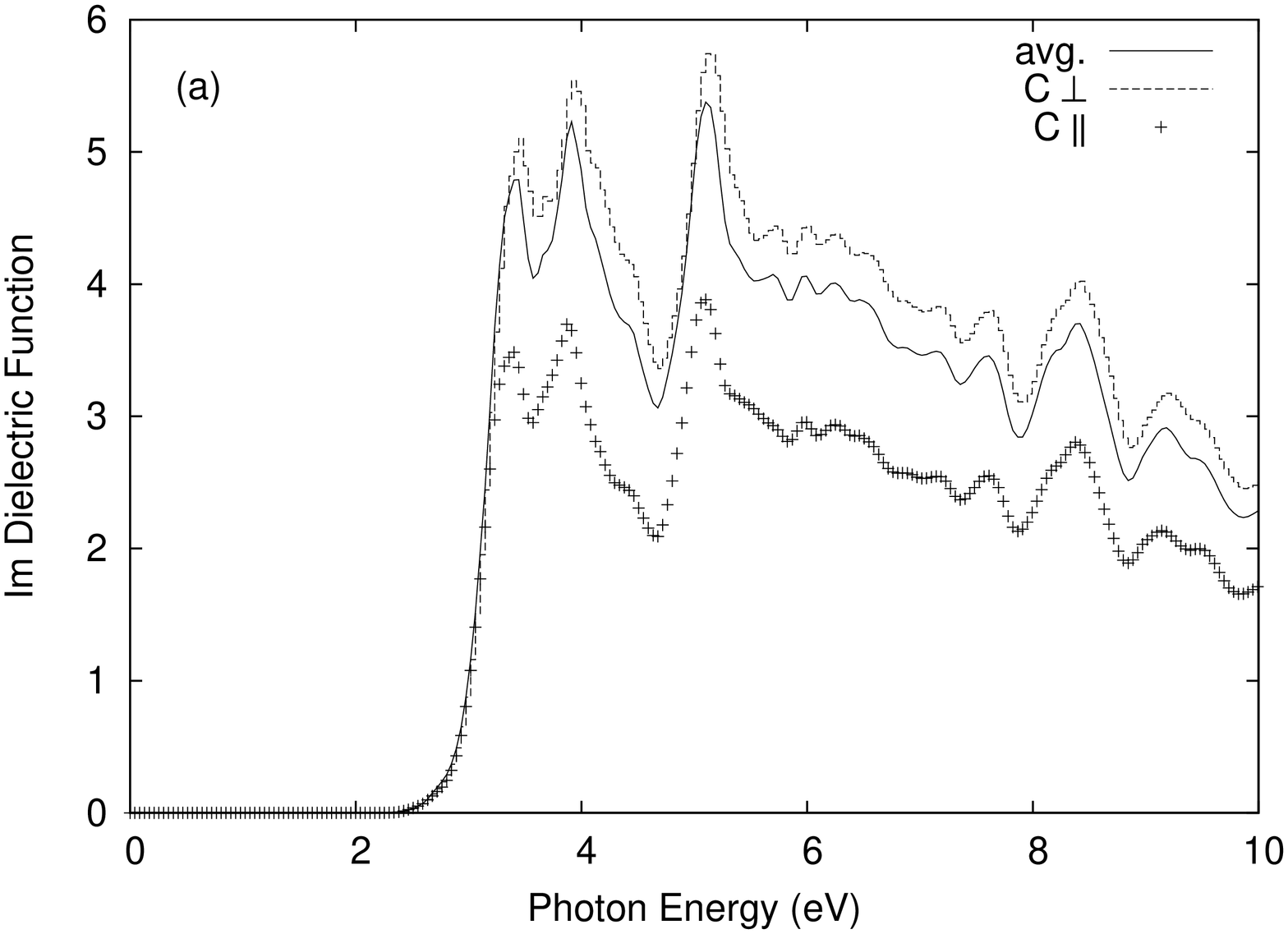}}}}} & {\rotatebox{-90}{\resizebox{7.0cm}{7.0cm}{\includegraphics{{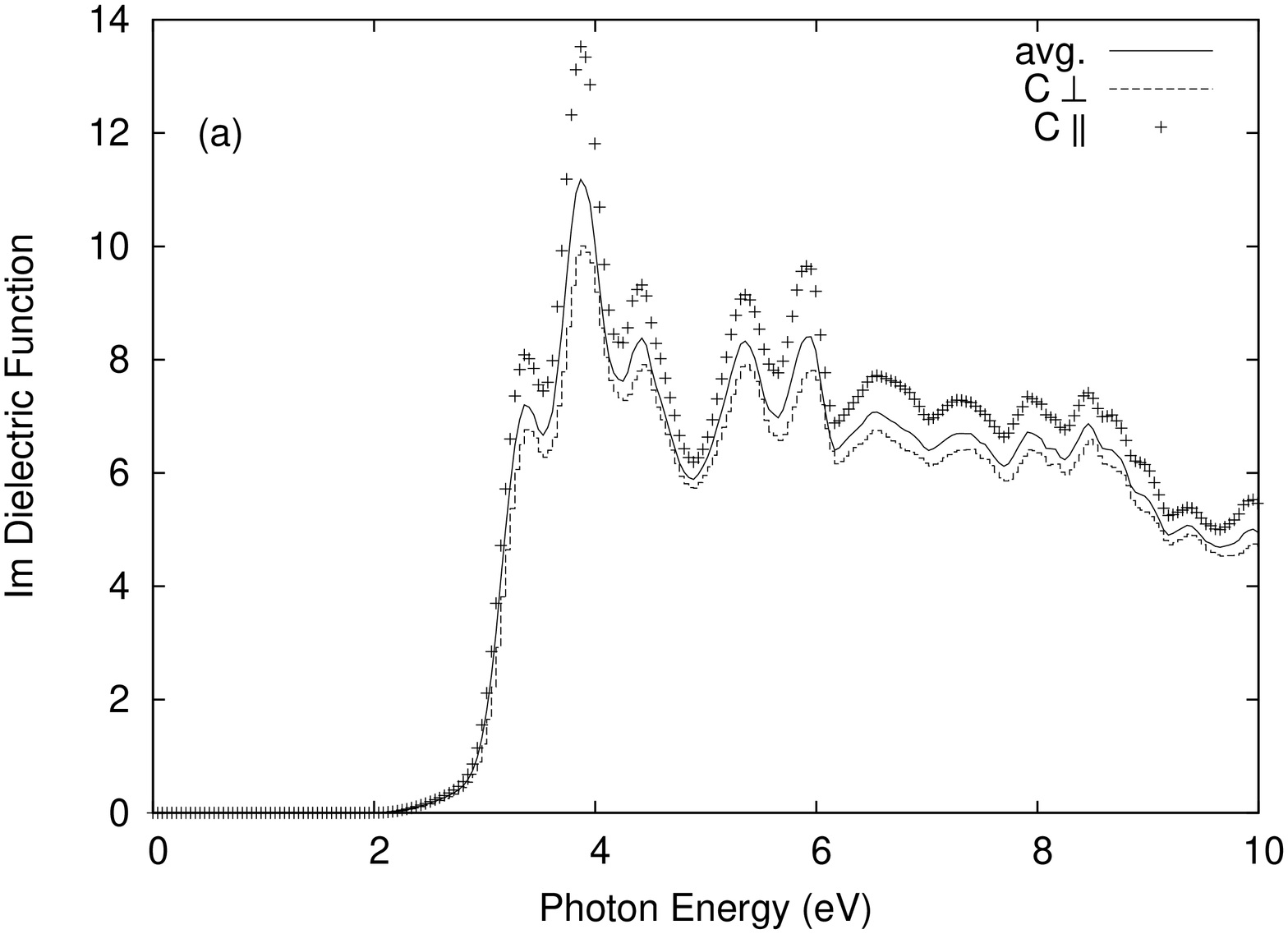}}}}}\\
{\rotatebox{-90}{\resizebox{7.0cm}{7.0cm}{\includegraphics{{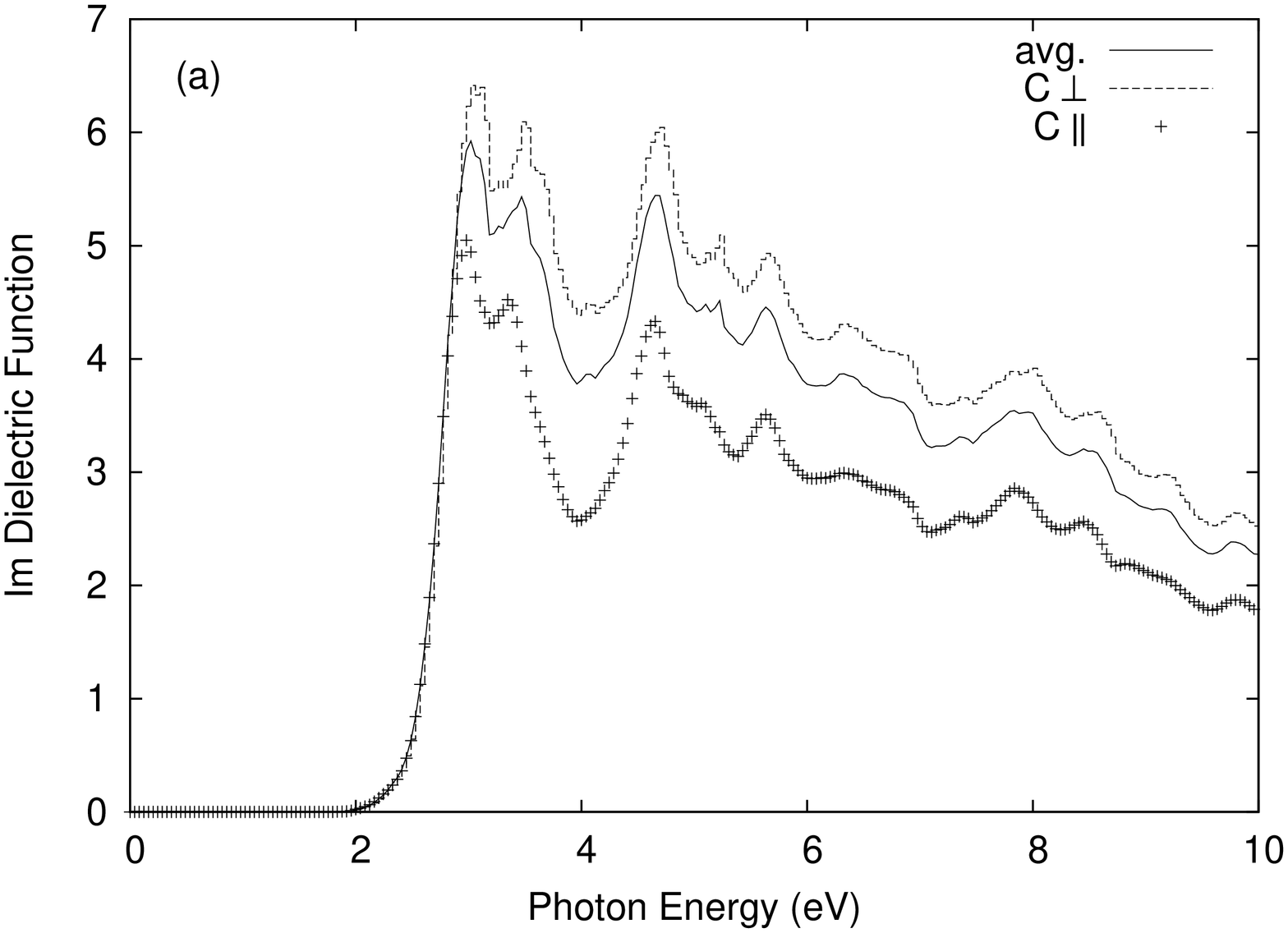}}}}} & {\rotatebox{-90}{\resizebox{7.0cm}{7.0cm}{\includegraphics{{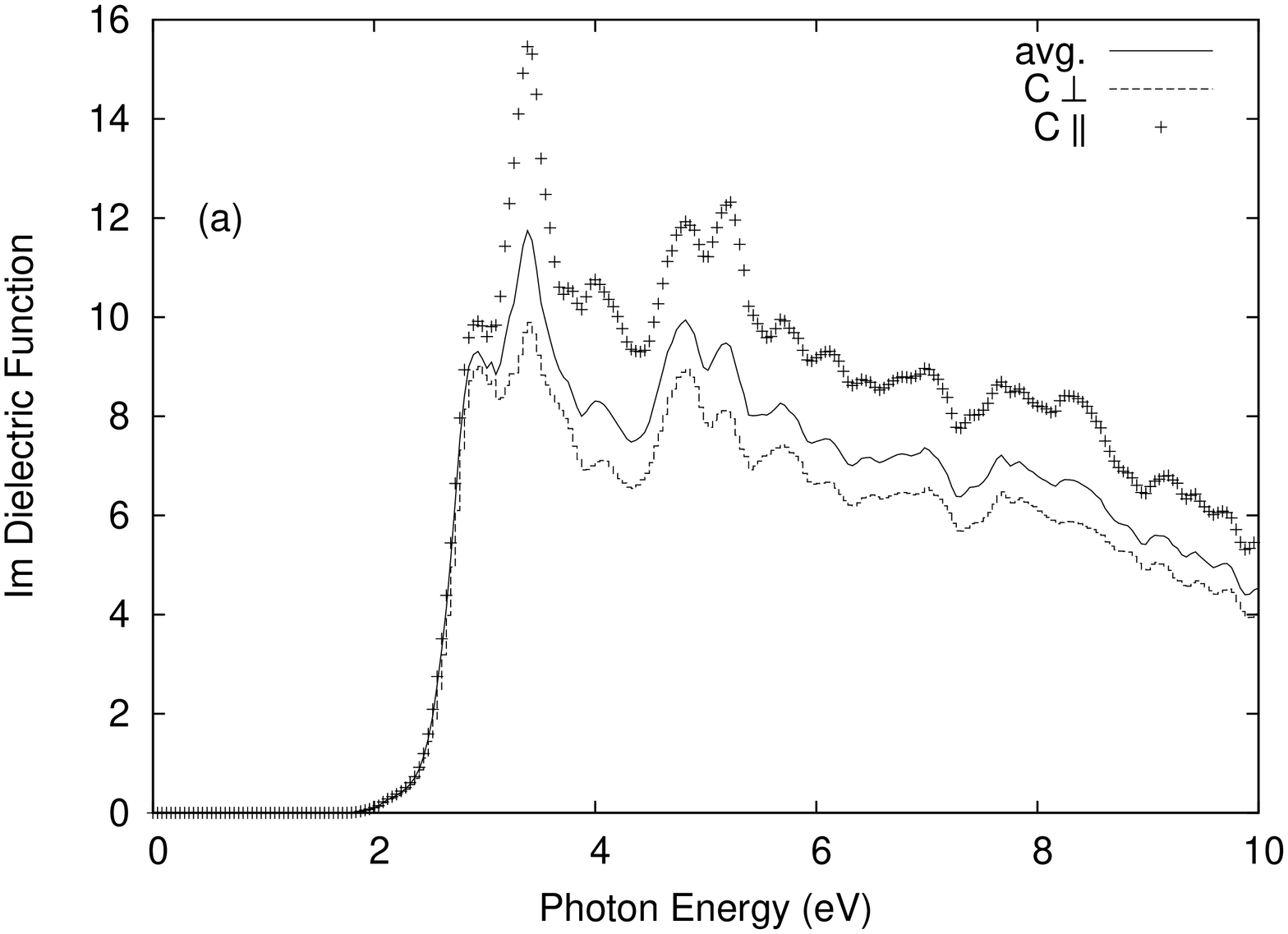}}}}}\\
{\rotatebox{-90}{\resizebox{7.0cm}{7.0cm}{\includegraphics{{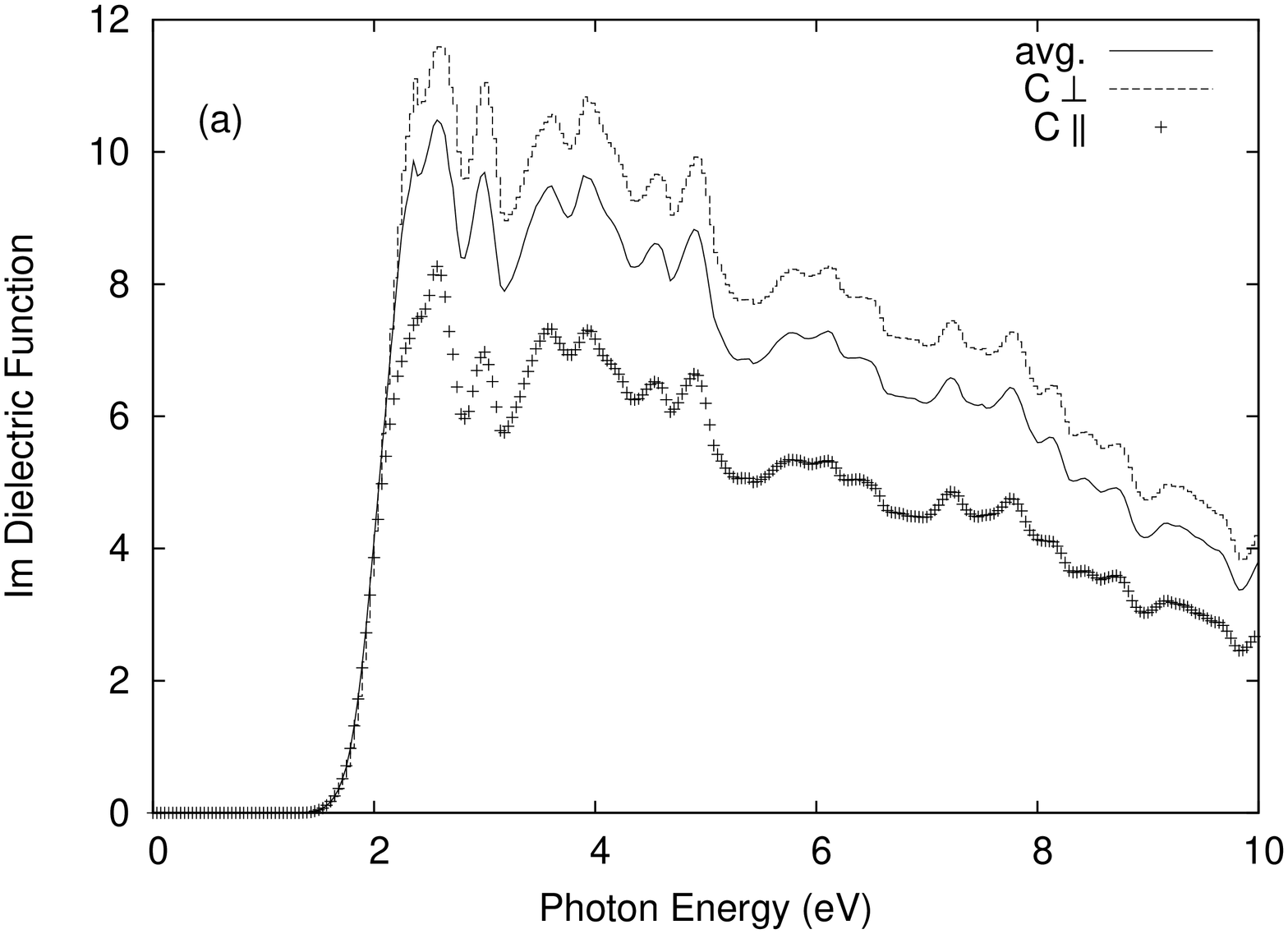}}}}} & {\rotatebox{-90}{\resizebox{7.0cm}{7.0cm}{\includegraphics{{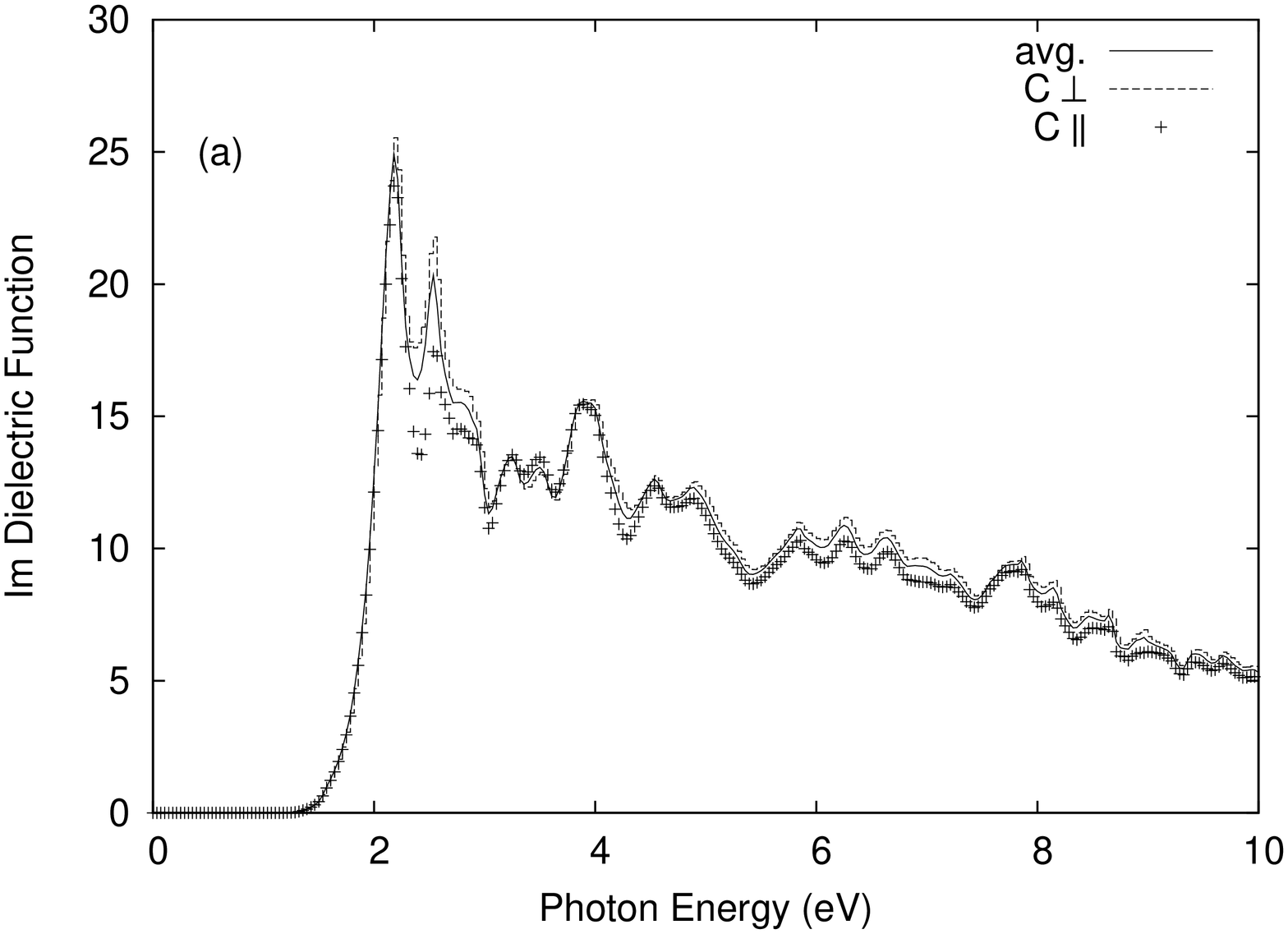}}}}}
\end{tabular}
\caption{(Left panel) : Imaginary part of dielectric function for non-ideal (right panel) : for ideal $CdGa_2X_4$ ; $X=S,Se,Te$ respectively}
\end{figure}

\begin{figure}
\centering
\begin{tabular}{cc}
{\rotatebox{-90}{\resizebox{7.0cm}{7.0cm}{\includegraphics{{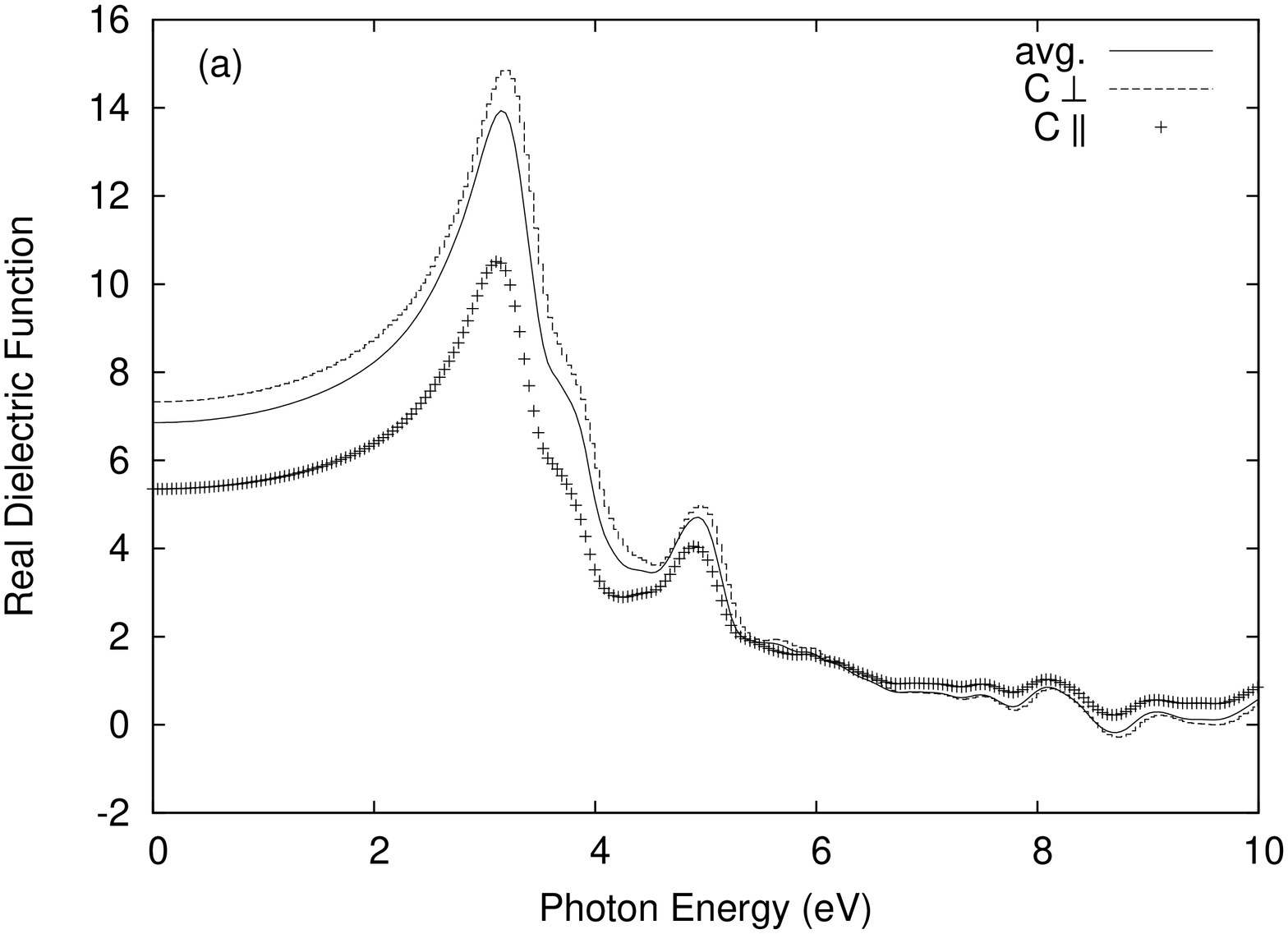}}}}} & {\rotatebox{-90}{\resizebox{7.0cm}{7.0cm}{\includegraphics{{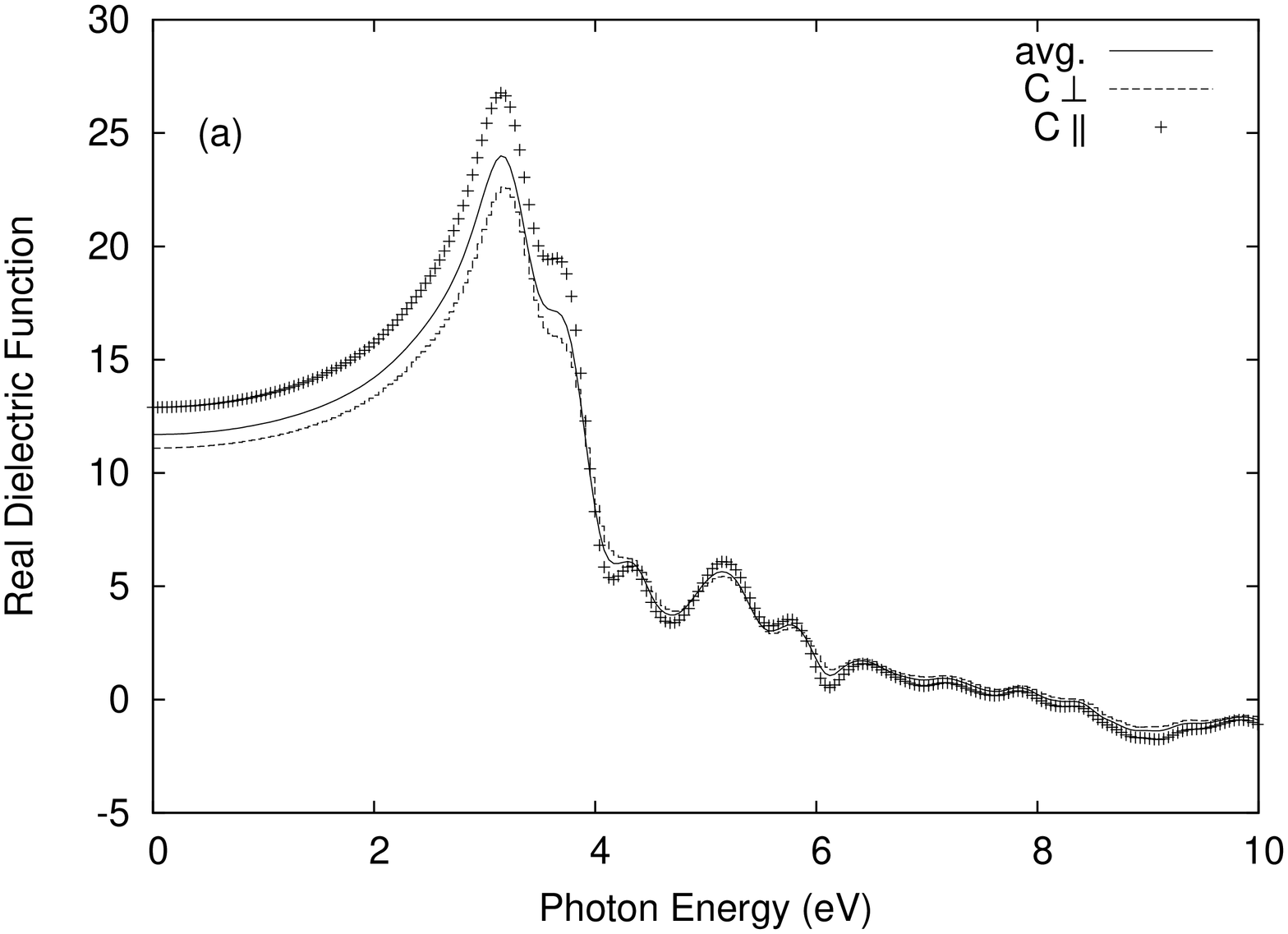}}}}}\\
{\rotatebox{-90}{\resizebox{7.0cm}{7.0cm}{\includegraphics{{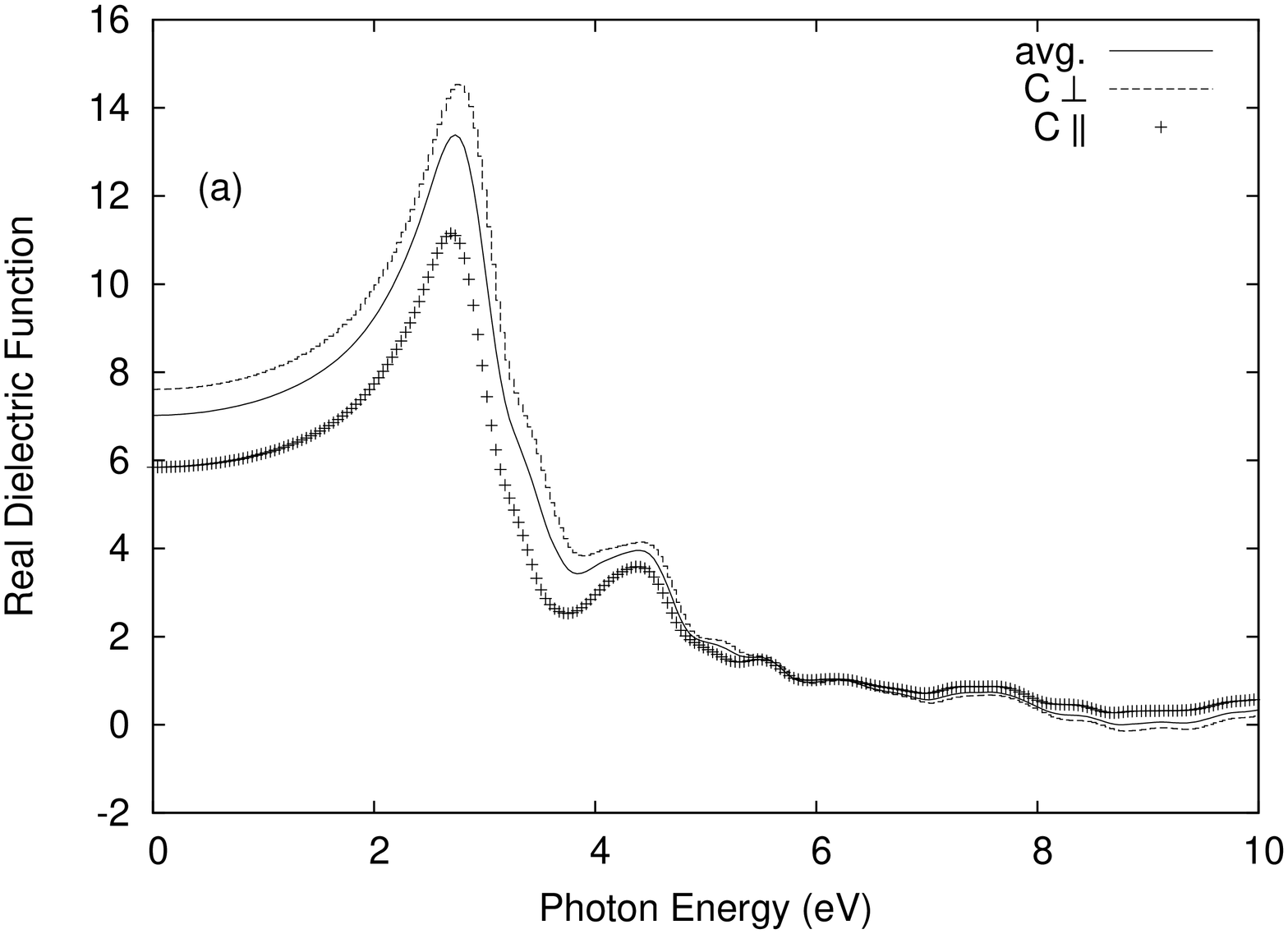}}}}} & {\rotatebox{-90}{\resizebox{7.0cm}{7.0cm}{\includegraphics{{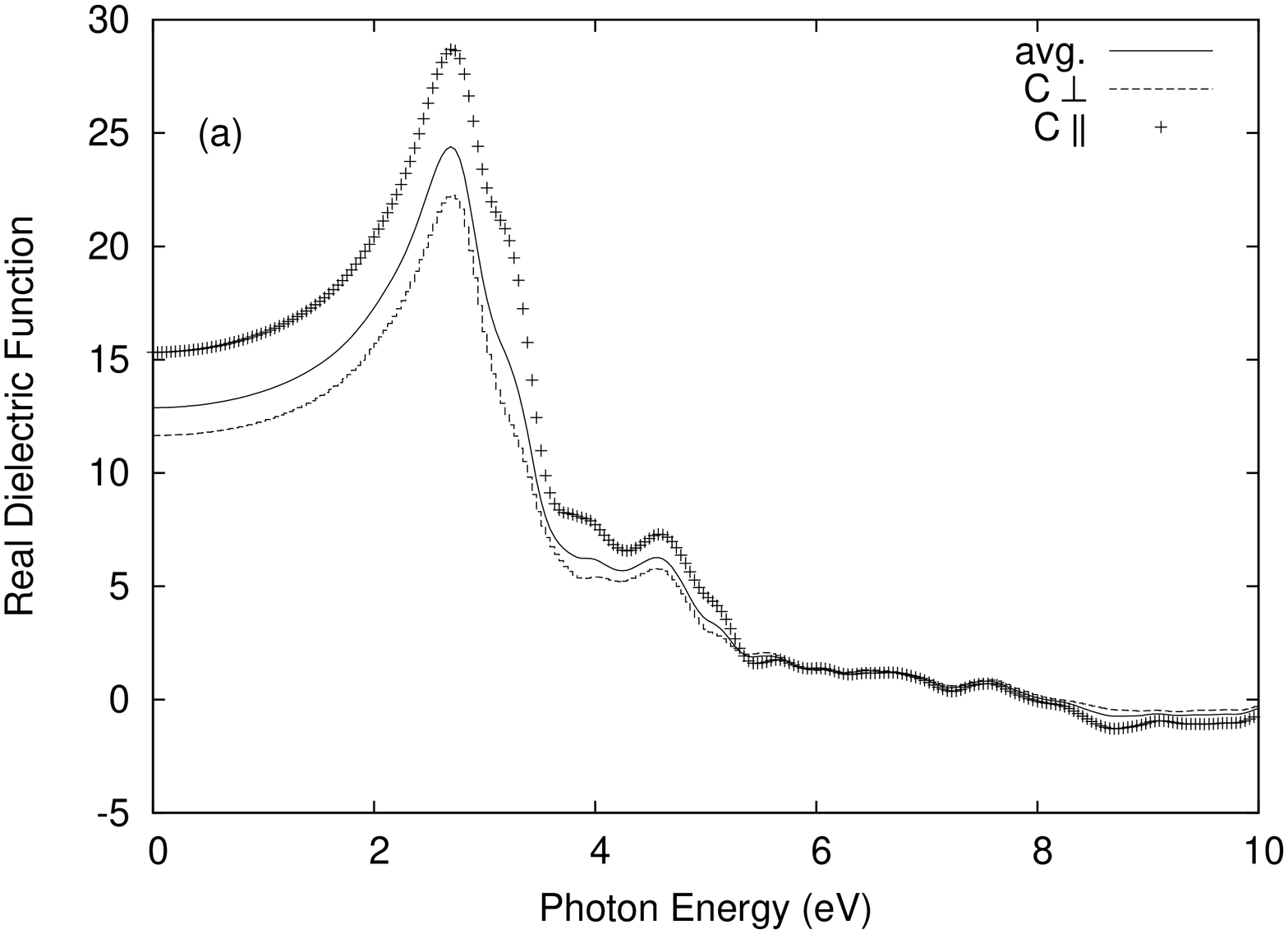}}}}}\\
{\rotatebox{-90}{\resizebox{7.0cm}{7.0cm}{\includegraphics{{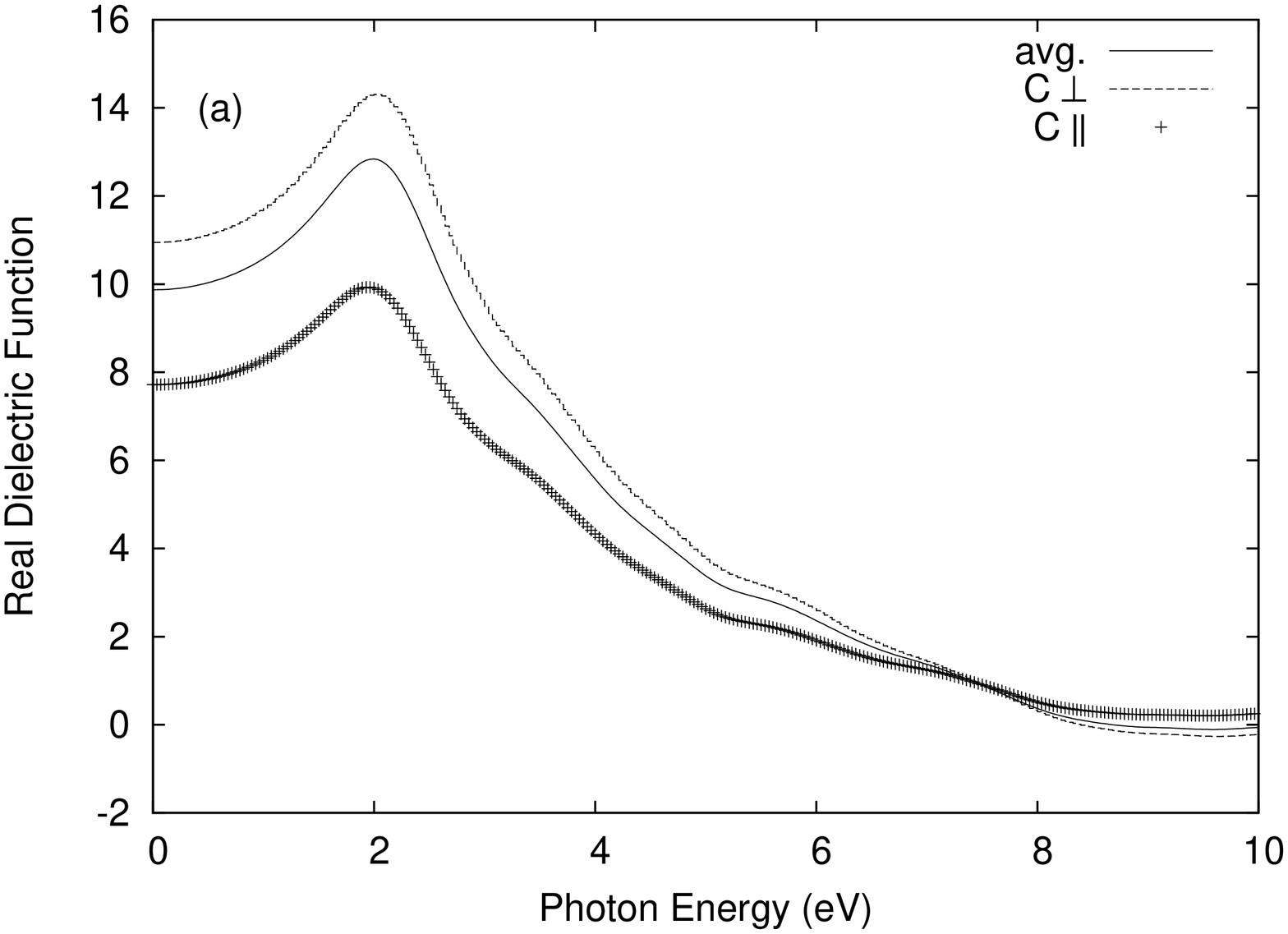}}}}} & {\rotatebox{-90}{\resizebox{7.0cm}{7.0cm}{\includegraphics{{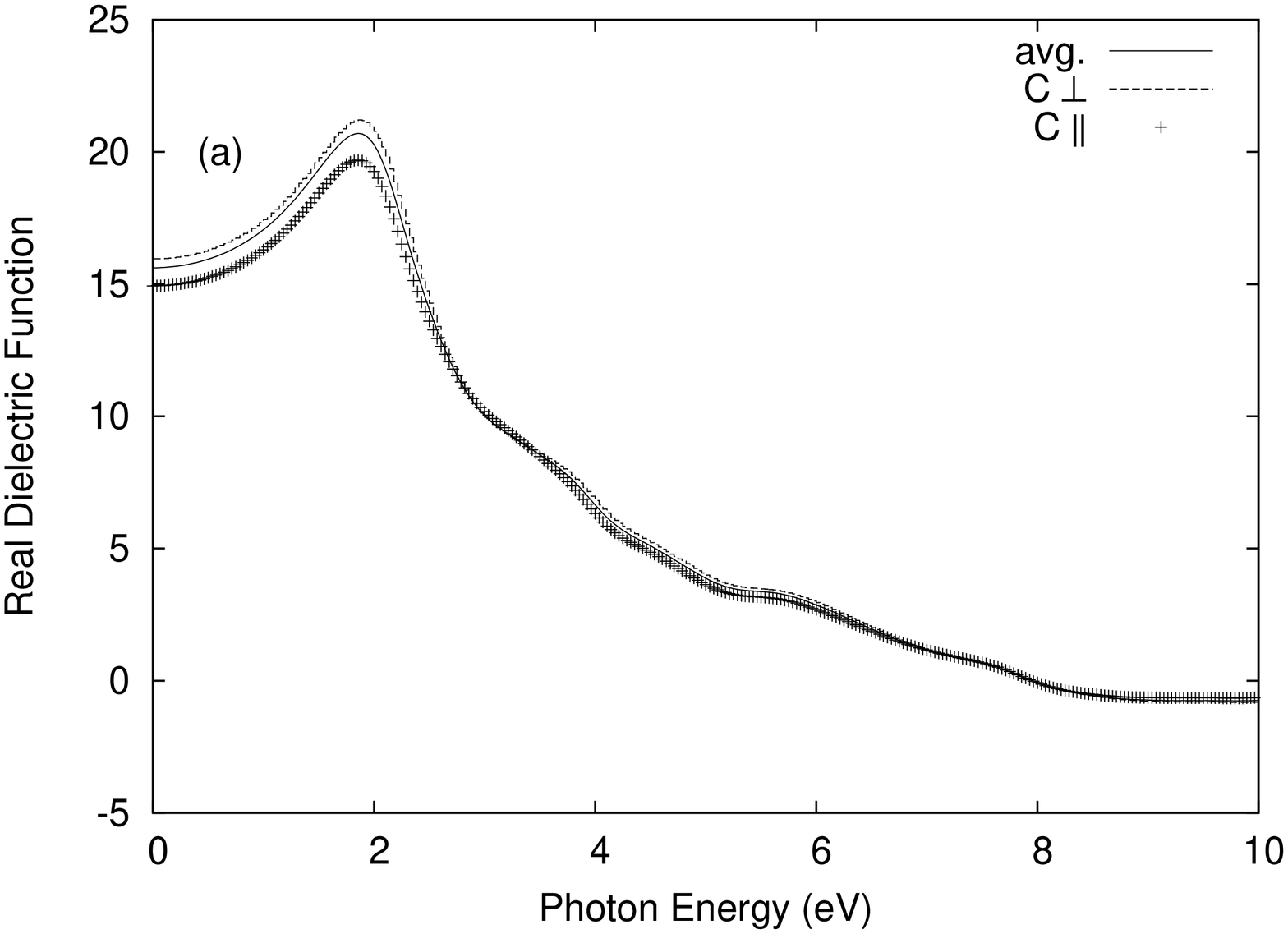}}}}}
\end{tabular}
\caption{(Left panel) : Real part of dielectric function for non-ideal (right panel) : for ideal $CdGa_2X_4$ ; $X=S,Se,Te$ respectively}
\end{figure}

\begin{table}
\begin{center}
\caption{Static dielectric constant $\epsilon_1(0)$ and refractive index $n(0)$. for non-ideal}
\begin{tabular}{cccc}
\hline
Optical Constants & $CdGa_2S_4$ &$CdGa_2Se_4$ & $CdGa_2Te_4$\\
\hline
$\epsilon_1(0)^{\perp}$ & 7.32  & 7.61 & 10.94  \\
$\epsilon_1(0)^{||}$&5.35 & 5.84 &7.71\\
$\epsilon_1(0)_{other}^{\perp}$&4.29$^a$ & 5.56$^a$ &9.53$^b$\\
$\epsilon_1(0)_{other}^{||}$& 4.22$^a$& 5.45$^a$ & 9.56 $^b$\\
$\epsilon_1(0)_{expt}^{\perp}$& - & - & 12.04$^c$\\
$\epsilon_1(0)_{expt}^{||}$& - & - & 11.47$^c$ \\
$n(0)^{\perp}$& 2.70 & 2.75& 3.30\\
$n(0)^{||}$& 2.31 & 2.41& 2.77\\
$n(0)_{other}^{\perp}$& 2.07$^a$ & 2.35$^a$ & 3.09$^b$\\
$n(0)_{other}^{||}$& 2.05$^a$ & 2.33$^a$& 3.09$^b$\\
$n(0)_{expt}^{\perp}$&2.3$^d$ & - & -\\
$n(0)_{expt}^{||}$& - & - & -\\
\hline
\end{tabular}
\end{center}
$^a$ Ref.[14]; $^b$ Ref.[19]; $^c$ Ref.[43]; $^c$ Ref.[44]
\end{table}

\begin{table}
\begin{center}
\caption{Static dielectric constant $\epsilon_1(0)$ and refractive index $n(0)$ for ideal $CdGa_2X_4$.}
\begin{tabular}{lcccc}
\hline
System & $\epsilon_1(0)^{\perp}$ & $\epsilon_1(0)^{||}$ & $n(0)^{\perp}$ & $n(0)^{||}$  \\
\hline
$CdGa_2S_4$  &  11.09 & 12.89 & 3.33 & 3.59  \\
$CdGa_2Se_4$ &  11.65 & 15.32 & 3.41 &3.91 \\
$CdGa_2Te_4$&   15.95 & 14.96 & 3.99 & 3.86  \\
\hline
\end{tabular}
\end{center}
\end{table}

\section{Conclusion}
Calculations and study of structural, electronic and optical properties of $CdGa_2X_4 (X=S,Se,Te)$ suggest that all these three compounds are direct band gap semiconductors with band gap 2.20, 1.75 and 1.25 eV respectively. Our study further shows that electronic properties such as band structure, TDOS and PDOS of these semiconductors significantly depend on the structural distortion. All these calculations are carried out using DFT based TB-LMTO method. We use LDA for our exchange corelation functional. Taking into account the underestimation of band gap by LDA, our result of band gap and structural properties agree with the available experimental values. We find  an inecreament of band gaps by 3.63$\%$, 4.00$\%$ and 8.8$\%$ respectively for $CdGa_2X_4$ (S, Se, Te) due to structural distortion. We also find significant effects of structural distortion on optical properties of these compounds. We have calculated joint density of states (JDOS), square of optical matrix elements (OME) and the optical response functions like real and imaginary part of dielectric functions, static dielectric constants. We conclude that main effects of structural distortion on optical properties come from the significant changes in OME due to structural variations. Results of real and imaginary parts of dielectric constants, static dielectric constants of these compounds agree well with the available experimental results. 
\section*{Acknowledgement}
This work was supported by Department of Science and Technology, India, under the grant no.SR/S2/CMP-26/2007. We would like to thank Prof. O.K. Andersen, Max Planck Institute, Stuttgart, Germany, for kind permission to use the TB-LMTO code developed by his group.

\end{document}